\documentclass[11pt,a4paper]{article}
\pdfoutput=1 
\usepackage{jcappub}
\usepackage{amsmath}
\usepackage{amsfonts}
\usepackage{amssymb}
\usepackage{graphicx}
\usepackage[usenames,dvipsnames]{color}
\usepackage{hyperref}
\usepackage{mathrsfs}

\newcommand{\abs}[1]{\left\vert#1\right\vert}
\newcommand{\set}[1]{\left\{#1\right\}}

\newcommand{\eps}{\varepsilon}

\newcommand{\ord}[1]{\textrm{ord}\left(#1\right)}
\newcommand{\bund}{\mathscr{B}}
\newcommand{\pb}{\textrm{\tiny PB}}

\newcommand{\del}{\boldsymbol{\nabla}}

\def\actaa{\ref@jnl{Acta Astron.}}      

\title{Multiple-Scales Approach to The Averaging Problem in Cosmology}%
\author[a]{Yonadav Barry Ginat}%
\affiliation[a]{Faculty of Physics, Technion -- Israel Institute of Technology, Haifa, 3200003, Israel}%
\emailAdd{ginat@campus.technion.ac.il}%

\abstract{
  The Universe is homogeneous and isotropic on large scales, so on those scales it is usually modelled as a Friedmann-Lema\^{i}tre-Robertson-Walker (FLRW) space-time. The non-linearity of the Einstein field equations raises concern over averaging over small-scale deviations form homogeneity and isotropy, with possible implications on the applicability of the FLRW metric to the Universe, even on large scales. Here I present a technique, based on the multiple-scales method of singular perturbation theory, to handle the small-scale inhomogeneities consistently. I obtain a leading order effective Einstein equation for the large-scale space-time metric, which contains a back-reaction term. The derivation relies on a series of consistency conditions, that ensure that the growth of deviations from the large-scale space-time metric do not grow unboundedly; criteria for their satisfiability are discussed, and it is shown that they are indeed satisfied if matter is non-relativistic on small scales. The analysis is performed in harmonic gauge, and conversion to other gauges is discussed. I estimate the magnitude of the back-reaction term relative to the critical density of the Universe in the example of an NFW halo, and find it to be of the order of a few percent. In this example, the back-reaction term is interpreted as a contribution of the energy-density of gravitational potential energy, averaged over the small-scale, to the total energy-momentum tensor.
}

\begin{document}
\maketitle
\flushbottom

\section{Introduction}
In Newtonian mechanics, Newton's third law ensures that the internal forces acting inside a material body are immaterial to understanding how it moves -- \emph{that} is determined only by the forces exerted on it by other objects. A collection of particles may be viewed as a single object when `zooming out', and treated as a point particle on scales that are large enough (provided it does not break up). If there are no spatial boundaries (their existence impacts on integration by parts), there is no back-reaction in Newtonian gravity, as it is a linear theory \cite{BuchertEhlers1997}.

Einstein's equations are non-linear though, and therefore concern has risen that when one studies the large-scale structure of space-time, the homogeneity of matter on large scales does not imply that inhomogeneities on small scale do not influence the large-scale metric \citep{Clarksonetal2011,BuchertRaesaenen2012}. Indeed, the non-linearity of the Einstein equations
\begin{equation}\label{eqn:Einstein}
  R_{ab} - \Lambda g_{ab} = 8\pi G\rho_{ab},
\end{equation}
where
\begin{equation}
  \rho_{ab} = T_{ab} - \frac{1}{2}Tg_{ab},
\end{equation}
($T^{ab}$ is the energy-momentum tensor) implies that averaging over spatial scales cannot be done easily -- it does not commute with metric inversion, the connection, \emph{et cetera}. Considered as an initial value problem, the evolution of the averaged spatial metric due to the exact equations is not, in general, the same as the evolution of the metric generated by the averaged equation; this affects the extent to which the Friedmann-Lema\^{i}tre-Robertson-Walker (FLRW) solution is valid as a description of the Universe on large scales \cite{Buchert2000}. The difference arises from small-scale inhomogeneities, that react back on large scales through the non-linearity of the Einstein equations.
This problem has been studied a lot in recent years (see e.g. \cite{Fuatmase1996,Zalaletdinov1997,Buchert2000,Kolbetal2005,Wiltshire2008,Gasperinietal2011,GreenWald2011,Raesaenen2012,Korzynski2015,BentivegnaBruni2016,Goldbergetal2017,Callagheretal2019}), but the magnitude of this so-called `back-reaction' and of its influence on the large-scale gravitational dynamics of the Universe is still a matter of some debate \citep{GreenWald2014,Buchertetal2015,GreenWald2016}. Some numerical relativistic simulations were conducted to investigate the averaging problem \citep{BentivegnaBruni2016,Adameketal2018,Macphersonetal2019}, leading to the conclusion that the over-all effect is probably small, and depends on the space-time slicing.

It is clear that standard cosmological perturbation theory does not suffice to handle the back-reaction problem \cite{EllisColey2019}, and a novel technique is needed -- be it an averaging technique \cite{Buchert2000,Gasperinietal2011}, or a special asymptotic expansion \cite{GreenWald2011,Goldbergetal2017,Callagheretal2019}. Here, I wish to propose such an approach, which utilises the multiple-scales method of singular perturbation theory, explained below in \S \ref{subsec: mutliple-scales exposition}.

I use harmonic gauge throughout the paper, and, of course, some of the results may be gauge-dependent. In harmonic co-ordinates equations \eqref{eqn:Einstein} are quasi-linear hyperbolic equations, with the Ricci tensor given by \cite{Choquet-Bruhat2009}
\begin{equation}\label{eqn:Ricci harmonic}
  R_{ab} \equiv R^{(h)}_{ab} = -\frac{1}{2}g^{cd}\partial^2_{cd}g_{ab} + P^{cdefgh}_{ab}(g)\partial_c g_{ef}\partial_d g_{gh},
\end{equation}
where
\begin{equation}\label{eqn:P_abcdefgh}
  P^{cdefgh}_{ab}(g)\partial_c g_{ef}\partial_d g_{gh} = -\frac{1}{2}\left(\partial_b g^{cd}\partial_c g_{ad} + \partial_a g^{cd}\partial_c g_{bd}\right) - \Gamma^c_{ad}\Gamma^d_{bc}.
\end{equation}
When it does not cause confusion, I shorten $P^{cdefgh}_{ab}(g)\partial_c g_{ef}\partial_d g_{gh}$ to $P_{ab}(g)\partial g\partial g$. I use the following order notation: $g(\eps) = O(f(\eps))$ if $\lim_{\eps \to 0} \abs{g(\eps)/f(\eps)}$ is bounded, $g(\eps) = o(f(\eps))$ if $g(\eps)/f(\eps) \underset{\eps \to 0}{\to} 0$, and $g(\eps) = \ord{f(\eps)}$ if $g(\eps)/f(\eps) \underset{\eps \to 0}{\to} \textrm{const} \neq 0$. To avoid some complications, I take space-time to be globally-hyperbolic.

\subsection{The Method of Multiple Scales}
\label{subsec: mutliple-scales exposition}
The multiple-scales method \citep{Jikovetal1994,KevorkianCole1996,Johnson2005,PavliotisStuart2008} has wide-ranging applications throughout physics \cite{BenderBettencourt1996,BattyeMennim2004,Mengetal2019}. For instance, the Chapman-Enskog expansion, used in deriving the Navier-Stokes equations from the Boltzmann equation relies on it (see, e.g., \citep{Dellar2007}), as well as any homogenisation technique used to study diffusion or transport processes in inhomogeneous media \cite{PavliotisStuart2008}.

To introduce this method, and how it can handle small-scale effects consistently, let me review the multiple-scales homogenisation of a simple steady-state diffusion equation in one dimension (this can be found in many textbooks, e.g. \cite[chapter 12]{PavliotisStuart2008}). Consider, then, the differential equation
\begin{equation}\label{eqn: diffusion}
  \frac{\mathrm{d}}{\mathrm{d}x}\left(D\left(x,\frac{x}{\eps}\right) \frac{\mathrm{d}u}{\mathrm{d}x}\right) = 0,
\end{equation}
with unknown $u(x)$, where $D(x,y)$ is a known diffusion coefficient, bounded from below and from above by two positive constants, and where $\eps \ll 1$. $D(x,x/\eps)$ may have some small-scale oscillations, and one is interested in the large-scale behaviour of $u(x)$.

The method of multiple scales enables one to derive an asymptotic expansion $u \sim u_0 + \eps u_1 + \ldots$, such that $u_0$ is independent of $\eps$, and satisfies a so-called \emph{homogenised} equation, with a homogenised diffusion coefficient, $D_H(x)$, that depends on $D$. This is done by introducing a new, fast variable $X = x/\eps$, and treating $x$ and $X$ as \emph{independent variables}, and changing
\begin{equation}
  \frac{\mathrm{d}}{\mathrm{d}x} \mapsto \frac{\partial}{\partial x} + \frac{1}{\eps} \frac{\partial}{\partial X}.
\end{equation}
Treating $X$ and $x$ independently is akin to taking a function $f(z,w)$, with $w = \alpha z$, and then identifying the total derivative $\frac{\mathrm{d}f}{\mathrm{d}z}$ with
\begin{equation}
  \frac{\partial f}{\partial z} + \alpha\frac{\partial f}{\partial w},
\end{equation}
but going in the opposite direction.

This substitution turns equation \eqref{eqn: diffusion} into
\begin{equation}\label{eqn: diffusion scales}
  \eps\partial_x[D(\eps\partial_x u + \partial_X u)] + \partial_X[D(\eps\partial_x u + \partial_X u)] = 0.
\end{equation}
The fact that $D$ depends on $x$ and $X$ implies that $u = u(x,X)$, too. The next step is to postulate an asymptotic expansion for $u(x,X)$ of the form $u \sim u_0(x,X) + \eps u_1(x,X) + \eps^2 u_2(x,X) + \ldots$, substitute into equation \eqref{eqn: diffusion scales}, and solve order-by-order in $\eps$. The zeroth-order equation is $\partial_X[D~\partial_X u_0] = 0$, which implies that
\begin{equation}
  u_0 = b(x) + \int_0^X\frac{a(x)\mathrm{d}X'}{D(x,X')},
\end{equation}
for some functions $a,b$. Assuming that $u_0$ is bounded implies $a(x) = 0$ (because $D$ is bounded from both sides by positive constants), whence $u_0(x,X) = b(x)$, i.e. $u_0$ is independent of $x$. The next order is
\begin{equation}
  \partial_X[D(\partial_x u_0 + \partial_X u_1)] + \partial_x[D~\partial_X u_1] = 0,
\end{equation}
whence there exist functions $c(x),d(x)$, such that
\begin{equation}\label{eqn: diffusion scales 1 solution}
  u_1 = d(x) - X\partial_x u_0 + c(x)\int_0^X\frac{\mathrm{d}X'}{D(x,X')}.
\end{equation}
For the asymptotic expansion of $u$ to be consistent, one must require that $\eps u_1$ be smaller than $u_0$. Thus, one must require a \emph{consistency condition}: that the two last terms in equation \eqref{eqn: diffusion scales 1 solution} vanish together, in the limit $X \to \infty$. Therefore, one can define a function
\begin{equation}
  D_H(x) = \lim_{X\to \infty} \frac{X}{\int_{0}^{X}\frac{\mathrm{d}X'}{D(x,X')}},
\end{equation}
which satisfies $c(x) = D_H(x) \partial_x u_0$.

The second-order equation is
\begin{equation}
  \partial_X[D(\partial_X u_2 + \partial_x u_1)] =-\partial_x[D(\partial_X u_1 + \partial_x u_0)] = -c'(x).
\end{equation}
Solving it yields
\begin{equation}
  u_2(x,X) = e(x)\int_{0}^{X}\frac{\mathrm{d}X'}{D(x,X')} - c'(x)\int_{0}^{X}\frac{X'\mathrm{d}X'}{D(x,X')} - \int_{0}^{X}\partial_x u_1(x,X') \mathrm{d}X' + f(x).
\end{equation}
The first and the third terms are $\ord{X}$, while the second is $\ord{X^2}$ as $X\to \infty$, which necessitates a consistency condition that the coefficient of the second term must vanish, i.e., it must be the case that $c'(x) = 0$. Explicitly,
\begin{equation}\label{eqn: diffusion homogenised}
  \frac{\partial}{\partial x}\left[D_H(x) \frac{\partial u_0}{\partial x}\right] = 0.
\end{equation}
Equation \eqref{eqn: diffusion homogenised} is a diffusion equation for the large-scale, leading-order behaviour of $u$ -- the \emph{homogenised equation}, which has a non-trivial, homogenised diffusion coefficient, $D_H(x)$. Instead of a normal average, it is given by a harmonic average of the original diffusion coefficient over the small scale. It was derived by expanding the original equation in the two scales, and requiring, in each order in $\eps$, that the asymptotic expansion remain consistent.

\subsection{Outline}
The paper is structured as follows: I start by listing the assumptions on the energy-momentum tensor that are made in this paper, in \S \ref{sec:energy-momentum tensor}. These amount to, primarily, assuming that there exists a small number $\eps$, such that the energy-momentum tensor may be written as $\rho_{ab} = \rho_{ab}\left(x,X\right)$, where the $X \equiv x/\eps$ dependence pertains to small-scale structures, such as galaxies.

Then, using the multiple-scales method, I expand the Einstein equations in $\eps$ in \S \ref{sec:small-scale}; this implies that the partial derivatives split as
\begin{equation}
  \frac{\partial}{\partial x^a} \mapsto \frac{\partial}{\partial x^a} + \frac{1}{\eps}\frac{\partial}{\partial X^a}.
\end{equation}
The multiple-scales analysis also entails writing an asymptotic expansion
$g_{ab}(x,X) \sim g^0_{ab} + \eps g^1_{ab} + \ldots$,
as well as a similar expansion of the energy-momentum tensor. Assuming that the energy-momentum tensor is only due to Newtonian sources (see \S \ref{sec:Newtonian objects}), or that it is only $O(1)$ in $\eps$ (\S \ref{sec:small-scale}), the zeroth-order equation is shown to give rise to a metric $g^0_{ab}$ which is independent of $X$, and depends only on the large scale $x$.

The higher-order equations turn out to be wave equations with sources, whose wave operator is of the form $g_0^{cd}\partial^2_{X^cX^d}$. As in \S \ref{subsec: mutliple-scales exposition}, the consistency of the asymptotic expansion implies certain consistency conditions on the $g^1_{ab}$ \emph{etc.} -- in fact, this is a restriction on the source terms which are in the kernel of the wave operator -- that ensures that $g^1_{ab}$ \emph{etc.} remain bounded in $X$. The first non-trivial consistency condition appears only in the second order equation -- for $g^2_{ab}$ (\S \ref{subsec: oscillatory}). The consistency conditions in general are discussed in depth in \S \ref{subsec: consistency conditions stars}, explicitly up to $\ord{\eps^3}$, and their form at higher orders is also commented on. The primary purpose of the multiple-scales analysis in this paper is to show how these consistency conditions are satisfied, given the assumptions in \S \ref{sec:energy-momentum tensor}, and that this validates the asymptotic expansion, and also enables the derivation of an effective Einstein equation for $g^0_{ab}(x)$ in \S \ref{sec:Newtonian objects}, that contains correction terms, in which back-reaction may be manifested. This equation is of the form
\begin{equation}
  R_{ab} - \Lambda g_{ab} + B_{ab} = 8\pi G \langle \rho_{ab} \rangle,
\end{equation}
where $B_{ab}$ is defined in equation \eqref{eqn:back-reaction tensor} and $\langle \rho_{ab}\rangle$ -- in \S \ref{subsec:averaged part}. While the effective equation is similar to ones in the literature (e.g. those in refs. \citep{Zalaletdinov1997,GreenWald2011}), here, it rests on a different footing due to the explicit analysis of the consistency conditions.

In \S \ref{subsec:examples} three examples of space-times where the correction terms are estimated are considered: an NFW halo, a general virialised object, and an Einstein-Straus model. I also describe briefly a geometrical interpretation of the multiple-scale splitting in \S \ref{subsec:fibre bundles}. I finish with a discussion and conclusions in \S \ref{sec:discussion}.


\section{Assumptions Concerning The Energy-Momentum Tensor}
\label{sec:energy-momentum tensor}
The fundamental theoretical assumption that is made in this paper, is that matter in the Universe has a distribution that looks differently on different scales. These are comprised of stellar scales, galaxy-size scales, large-scale-structure (LSS) scales, and the largest, cosmological, scales \citep{Clarksonetal2011}. This implies that the energy-momentum tensor $\rho_{ab}$ describing the matter distribution in the exact solution of equation \eqref{eqn:Einstein} must naturally depend on all of those scales. If one measures distances on cosmological scales, in units of $100 ~\textrm{Mpc}$, then all the other scales, $\eps_{\textrm{stars}} \ll \eps_{\textrm{gal}} \ll \eps_{\textrm{LSS}}$, are much smaller than unity; they are also well-separated. Indeed, $\rho_{ab}$, as a function of position, must depend on it on all of these scales; its most general form is
\begin{equation}\label{eqn:general scale-dependent energy-momentum tensor}
  \rho_{ab} = \rho_{ab}\left(x,\frac{x}{\eps_{\textrm{stars}}},\frac{x}{\eps_{\textrm{gal}}} , \frac{x}{\eps_{\textrm{LSS}}}\right).
\end{equation}
One can always separate the dependence on $x$ into different scales in this way, at least formally: in Fourier space, one simply groups all modes whose wavelengths are smaller than $2\pi \eps_\textrm{stars} \times 100~\textrm{Mpc}$ into one group, then all the modes with wavelengths between $2\pi \eps_\textrm{stars}\times 100~\textrm{Mpc}$ and $2\pi \eps_\textrm{gal}\times 100~\textrm{Mpc}$ into another, \emph{et cetera}, while all the long modes are grouped into the cosmological-scales dependence of $\rho_{ab}$. In appendix \ref{appendix:filtering of rho} I propose one way, out of many, that one might use to find the dual $x,X$-dependence of $\rho_{ab}$ from a given tensor field $\rho_{ab}(x)$ on space-time.

In a wider context, a separation of scales in the matter distribution is common in the theory of large-scale structure, in the context of a peak-background split (see, e.g., \cite{Bardeenetal1986}), where it is used to describe the clustering of over-dense regions. It is also used in the study of gravitational waves, as a basis for a WKBJ expansion, aimed at deriving the geometric optical description of progressive gravitational waves (e.g. \cite{Isaacson1968a,Isaacson1968b}).

For simplicity, however, I will focus on matter on just two scales: galactic scales and cosmological scales. This will allow me to keep the introduction of the application of the multiple-scales method in this paper -- its main aim -- as simple as possible, while retaining the essential physics; thus, set $\eps = \eps_\textrm{gal}$, whence the assumption in equation \eqref{eqn:general scale-dependent energy-momentum tensor} becomes
\begin{equation}\label{eqn:energy-momentum}
  \rho_{ab} = \rho_{ab}\left(x,X\right),
\end{equation}
where $X = x/\eps$.
All modes with wavelengths smaller than $2\pi \eps_\textrm{gal}$ fall into the galactic-scales dependence -- on $X$ -- while all other modes are understood to contribute to the
$x$ dependence. In general, the metric would depend on $x$ in this way, too.

Equation \eqref{eqn:energy-momentum} is not begging the question -- writing $\rho_{ab}$ as a function of two variables, the large scale and the small scale, does not yield that there is no back-reaction. Rather, it is an \emph{observational statement} about the cosmological principle: the matter distribution in the universe depends on many scales, and its the energy-momentum tensor should do, too. In practical applications the way to define $x$ and $X$ is therefore observational, where the latter describes matter inside a galaxy, and the former -- the galaxy. Indeed, there are space-times with and without sizeable back-reaction whose energy-momentum tensor satisfies equation \eqref{eqn:energy-momentum} (e.g. \citep{WheelerLindquist1957,Kantowski1969,Cliftonetal2012,Bolejko2017}). Such a splitting into a small scale and a large one is part of the mathematical model that is used in this paper to address the averaging problem; form the point of view of the model, it is an assumption. From a physical point of view, on the other hand, the justification of this assumption rests on the existence of observational evidence for a scale-separation in the matter distribution in the Universe.

In what follows, I treat $\rho_{ab}$ as a known function of $(x,X)$, a source for the Einstein equations, rather than as a variable which both determines the gravitational field and is determined by it. This is merely a conceptual simplification that allows me to derive a leading order (in $\eps$) equation for the metric. Treating $\rho_{ab}$ as a source term in the Einstein equations means that the formalism developed here may be used to study the averaging problem as in a `post-processing' manner; that is, given an exact solution of the Einstein equations coupled to matter fields, one may ask if there is any back-reaction of small-scale variations in the matter distribution on the large-scale behaviour of the metric. I emphasise, though, that this is a conceptual change, rather than a restriction, because one may use it to determine if there is back-reaction in any specified solution.

$\rho_{ab}$ is only due to matter. However, when one makes cosmological observations to measure $\rho_{ab}$, the result is always dependent on the underlying metric. Observationally, it is impossible to disentangle the effect of the metric from a measurement of the matter content of the universe, without some other assumption, which might constitute some form of back-reaction; this is true, at least, for the large-scale behaviour of $\rho_{ab}$. The inferred matter density may be, for example, in part due to back-reaction of curvature terms, and have nothing to do with the number of particles in the Universe. However, if it were found, somehow, that the back-reaction term in the ``averaged'' Einstein equations are small, \emph{given} the true $\rho_{ab}$ (which is only due to matter), then it would follow \emph{a posteriori} that a measurement attempting to discover $\rho_{ab}$ would yield a tensor that is close to it.

If $x$ is measured in units where space starts to be homogeneous, $100$ Mpc, and as $\eps$ describes galactic scales of $1$ kpc, then $\eps = 10^{-5}$ (and then $X$ also has units of $100$ Mpc). Any $\ord{1}$ change in the value of $\eps$ does not affect the results below significantly. To summarise: the fundamental assumption on $\rho_{ab}$ is that (in harmonic co-ordinates) one could write $\rho_{ab} = \rho_{ab}(x,X)$ with $X = x/\eps$ and $\eps = 10^{-5} \ll 1$.

In the next section I start by ignoring over-densities in $\rho_{ab}$ that are larger than $\ord{1}$, as it is simpler to explain the method in this case; in \S \ref{sec:Newtonian objects} I include over-denisities due to galaxies. In doing so I assume that these small-scale over-densities in $\rho_{ab}$ are known to be only due to matter, and are observable directly. I still do not assume anything about the large-scale structure of the $O(1)$ component of $\rho_{ab}$ relative to the one inferred from observations, which is where back-reaction terms are important.

\section{Small-Scale Variations}
\label{sec:small-scale}
I use the method of multiple scales, treating $x$ and $X$ as \emph{independent} variables.
This procedure is valid as long as one can identify, observationally, different behaviours of the matter distribution on different physical scales. This amounts to requiring that only the short modes (with wave-length no more than $2\pi \eps\times 100~\textrm{Mpc}$) are allowed to have fluctuations of amplitude $\gg \ord{1}$. The smaller $\eps$ can be chosen such that it still satisfies this condition, the more accurate the asymptotic expansion derived in this section is. (Besides, if the intermediate-scale modes of $\rho_{ab}$ have a small, $o(1)$, amplitude relative to both cosmological and galactic scales, this sharpens the distinction between $x$ and $X$, and thus improves the accuracy of the approximation.)

\begin{figure}
  \centering
  \includegraphics[width=0.45\textwidth]{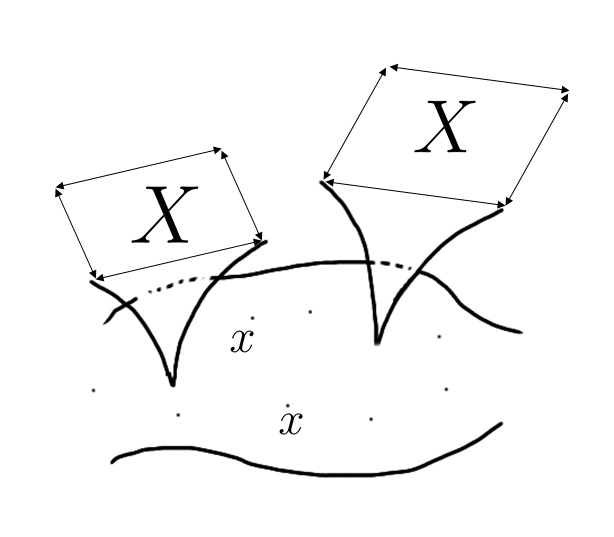}
  \caption{An illustration of the splitting of $x$ and $X$. The former describes the position of a galaxy in space-time, and the latter describes `zoomed-in' motion inside it.}\label{fig:illustration}
\end{figure}

The dimension of the manifold is then increased to $8$, and the partial derivative becomes
\begin{equation}\label{eqn:partial derivative splitting}
  \frac{\partial}{\partial x^a} \mapsto \frac{\partial}{\partial x^a} + \frac{1}{\eps}\frac{\partial}{\partial X^a}.
\end{equation}
In this view, $x$ specifies the position of a galaxy, whereas $X$ describes motion inside the galaxy (see figure \ref{fig:illustration}). From the point of view of the $x$-space-time galaxies are point particles, and $X$ `zooms into' each individual galaxy. The effect of anything that happens inside a given galaxy on the large scale emerges consistently from the coupling between $x$ and $X$ in the Einstein equations.

\subsection{Geometrical Interpretation}
\label{subsec:fibre bundles}
A geometrical interpretation of this construction may be provided in terms of fibre bundles. Let $M$ be space-time, and let $\bund = (M,F)$ be the bundle whose base is $M$, and whose fibres $F = \set{F_x}_{x\in M}$ are defined by the $X$-space-time. Thus, one inserts a new manifold, $F_x$, at every point $x \in M$. This might not be the trivial (product) bundle, as harmonic co-ordinates generally exist only locally. At present I do not specify the metric on $F$, but only require that it depend smoothly on $M$ (it will be shown later that $F_x$ may be treated as a flat space-time in \S \ref{subsec:low-order equations}, to zeroth order, in the scenarios I consider here). $F_x$ is a bounded manifold, with boundaries corresponding roughly to galaxy sizes. Local triviality follows from the equivalence principle.

The tensors I consider here are those appearing in equation \eqref{eqn:Einstein}, but they are (at present) defined only on the tangent space of $M$. However, if $T\bund$ is the tangent space of $\bund$, then it is locally spanned by
\begin{equation}
  \set{\frac{\partial}{\partial x^a},\frac{\partial}{\partial X^b}}_{a,b = 0,\ldots,3}.
\end{equation}
So, any tensor field $W^{ab}(x,x/\eps)$ on $M$ may be identified with a tensor field of the same rank on $\bund$, via
\begin{align}
  W & = W^{ab}\left(x,\frac{x}{\eps}\right)\frac{\partial}{\partial x^a}\otimes\frac{\partial}{\partial x^b} \\
  & \cong W^{ab}(x,X)\frac{\partial}{\partial x^a}\otimes\frac{\partial}{\partial x^b} + 0^{ab} \frac{\partial}{\partial X^a}\otimes \frac{\partial}{\partial X^b} + 0^{ab} \frac{\partial}{\partial x^a}\otimes \frac{\partial}{\partial X^b} + 0^{ab} \frac{\partial}{\partial X^b}\otimes\frac{\partial}{\partial x^b},
\end{align}
where $0^{ab}$ vanishes for each $a,b = 0,\ldots,3$. This means that $W$ is viewed as a tensor field on $\bund$, whose components depend smoothly both on $M$ and on $F$, but which really lies in the vector subspace of $T^{\otimes 2}\bund$ corresponding to $T^{\otimes 2}M$. Contrary to equation \eqref{eqn:partial derivative splitting}, the basis vectors do not split as the partial derivatives there -- that equation is relevant only for computing the components of $W^{ab}$.

\subsection{Asymptotic Expansion}
\label{subsec:asymptotic expansion}
I take the Universe to be well-described by standard cosmological perturbation theory up to recombination, which I set as the initial data for the multiple-scales calculation. The initial potential fluctuation (at recombination, i.e. deep in matter-domination) has a power spectrum
\begin{equation}\label{eqn:Power spectrum}
  P(k) = \frac{18\pi^2}{25}A_s \frac{k^{n_s-4}}{k_p^{n_s-1}}D^2(a_\textrm{rec})T^2(k),
\end{equation}
where $D(a)$ is the linear growth factor, $n_s \approx 1$, $A_s = 2.1\times10^{-9}$ \cite{Planck2018}, the pivot scale is $k_{p} = 0.05 ~\textrm{Mpc}^{-1}$, and the transfer function is $T(k) \approx \frac{12k_p^2}{k^2}\frac{k_\textrm{eq}^2}{k_p^2}\ln\left(\frac{k}{k_p}\right)$ \cite{Dodelson2003}, where $k_\textrm{eq} \approx 0.01 ~\textrm{Mpc}^{-1}$ \cite{Planck2018}. (This power spectrum is strictly valid for small scales only, which are precisely the scales I need it for.) This shape of the power-spectrum implies that the mean squared amplitude of a small-scale metric perturbation is
\begin{equation}
  \sigma_{\eps}^2 \propto \int_{k_0/\eps}^{\infty} \mathrm{d}k \frac{k^{n_s+2}\ln^2(k)}{k^8} \approx \int_{k_0/\eps}^{\infty} \mathrm{d}k \frac{\ln^2(k)}{k^5} = \frac{8\ln^2(k_0/\eps) + 4\ln(k_0/\eps) + 1}{32 (k_0/\eps)^4},
\end{equation}
where $k_0$ is some order-unity wave-vector whose exact value is immaterial. For $\eps = 10^{-5}$, $\ln^2\eps = \ord{1}$, whence $\sigma_{\eps}^2 = \ord{\eps^4}$, and the initial small scale (i.e. large $k$) metric perturbations have a root-mean-square amplitude proportional\footnote{The proportionality coefficient
  $\sqrt{\frac{2592\pi^2}{25}A_sk_p^{5-n_s}D^2(a_\textrm{rec})}$
is about $10^{-3}$, which might lead one to add an additional power of $\sqrt{\eps}$ to $\sigma_{\eps}$, making it $\ord{\eps^{5/2}}$. Doing this does not make any difference to what follows, so, for the sake of generality, I still include $\sigma_{\eps}$ in the $\ord{\eps^2}$ equations below, as a worst-case possibility. This also simplifies the expansion, relieving one of the need to expand in powers of $\sqrt{\eps}$. Naturally, different initial power spectra might, in general, require different expansions in $\eps$.} to $\eps^2$. The initial conditions for small-scale metric derivatives are, in virtue of the splitting of $x$ and $X$, $\partial_X g = O(\eps^2)$, because the derivative removes one power of $\eps$, which is restored by the relation $X = x/\eps$.

Thus, it is reasonable to expand in integer powers of $\eps$ in the asymptotic expansion of the metric and the energy-momentum tensor. Explicitly,
\begin{align}
   g_{ab}(x,X) & \sim g^0_{ab}(x,X) + \eps g^1_{ab}(x,X) + \eps^2 g^2_{ab}(x,X) + \mbox{h.o.t.} \label{eqn:asymptotic expansion metric}\\
   g^{ab}(x,X) & \sim g_0^{ab}(x,X) + \eps g_1^{ab}(x,X) + \eps^2 g_2^{ab}(x,X) + \mbox{h.o.t.} \\
   \rho_{ab}(x,X) & \sim \rho^0_{ab}(x,X) + \mbox{h.o.t.}.
\end{align}
For consistency, the various terms in this expansion would have to remain bounded, so that the hierarchy of orders is preserved throughout the system's evolution. I do not use any higher order terms in the expansion of $\rho_{ab}$ explicitly in this paper, so in fact, it is possible to expand it in different powers of $\eps$ from those in the series expansion of $g_{ab}$. Of course, the asymptotic series for $g^{ab}$ is determined in terms of that of $g_{ab}$ completely, to ensure that $g^{ab}$ is indeed the inverse of $g_{ab}$, at each order; for example $g_1^{ab} = -g_0^{ac}g_0^{bd}g^1_{cd}$. The reader should bear in mind that so far $g^0_{ab}$ is a completely general tensor-valued function of both $x$ and $X$, and may differ from an FLRW metric considerably. The expansion in equation \eqref{eqn:asymptotic expansion metric} is simply an expansion of a general function $g_{ab}(x,X,\eps)$.

In this expansion, the second-order differential operator in equation \eqref{eqn:Ricci harmonic} becomes
\begin{equation}\label{eqn:wave operator}
\begin{aligned}
  & -\frac{1}{2\eps^2}\bigg[\left(g_0^{cd}(x,X) + \eps g_1^{cd}(x,X) + \eps^2 g_2^{cd}(x,X)\right)\times \\ &
  \left(\eps^2 \partial^2_{x^cx^d} + 2\eps\partial^2_{x^cX^d} + \partial^2_{X^cX^d}\right)\times \\ &
  \left(g^0_{ab}(x,X) + \eps g^1_{ab}(x,X) + \eps^2 g^2_{ab}(x,X)\right)\bigg] + \ldots,
\end{aligned}
\end{equation}
while the first-order differential term reads
\begin{equation}\label{eqn:first differential operator}
\begin{aligned}
  & \frac{1}{\eps^2}\bigg\{ P_{ab}(g_0 + \eps g_1 + \eps^2 g_2)\times \\ &
  \left(\eps\partial_x + \partial_X\right)(g_0 + \eps g_1 + \eps^2 g_2) \times\\ &
  \left(\eps\partial_x + \partial_X\right)(g_0 + \eps g_1 + \eps^2 g_2)\bigg\} + \ldots.
\end{aligned}
\end{equation}

Let me remind the reader that in this section, I assume that $8\pi G\rho_{ab}$ is $O(1)$, at most (in \S \ref{sec:Newtonian objects} I relax this assumption), to be able to describe the method more easily, without the complications arising from a large energy-momentum tensor. To progress, I multiply the Einstein equations by $\eps^2$, whence the $\Lambda g_{ab}$ and $ 8\pi G\rho_{ab}$ terms contribute only at second order.\footnote{Even in generalised harmonic co-ordinates \cite{Choquet-Bruhat2009}, the additional contribution to the Ricci tensor is $O(\eps^2)$.} The zeroth order equation is
\begin{equation}\label{eqn:zeroth order}
  -\frac{1}{2}g^{cd}_0\partial^2_{X^cX^d}g^0_{ab} + P_{ab}(g_0)\partial_Xg_0\partial_Xg_0 = 0.
\end{equation}
This is a vacuum Einstein equation with no cosmological constant on $F_x$. Let me present a simple argument why this equation implies that $\partial_X g_0 = 0$, which is valid for $X$-independent initial data:\footnote{The general case can be handled with the results of \S \ref{subsec: consistency conditions stars}.} if the initial conditions are such that there are no order unity small-scale contributions to the metric, $g_0(\cdot,X)$ satisfies a vacuum Einstein equation, with constant initial conditions (i.e. flat space), whence by uniqueness, $g_0(\cdot,X)$ is independent of $X$. This is not a \emph{petitio principii}, for the only initial conditions used are at (say) recombination. Then, there are no zeroth-order small-scale perturbations to the metric, whence, at any $x$ such that $t = t_\textrm{rec}$, $\partial_X g_0(x,X) = 0$, even when $X^0$ reaches its maximum value. Hence, at a slightly later time $t = t_\textrm{rec} + \delta t$, $\partial_X g_0 = 0$ (in effect, one has a matching condition to ensure that the small scale behaviour does correspond to $X = x/\eps$). At this new (slow) time, one solves equation \eqref{eqn:zeroth order} again, giving the same result. The final consequence of this analysis is, that if there are $X$-independent initial conditions for a function $f(x,X)$ at recombination, and if the differential equation satisfied by $f$ (with respect to $X$ -- the large-scale coordinate $x$ is treated as a parameter) is such that $\partial_X f$ remains zero as a function of $X$, then $\partial_X f(x,X) = 0$, even for later times $t$.

\subsection{Low-Order Equations}
\label{subsec:low-order equations}
The order $\eps$ equation is
\begin{equation}\label{eqn:first order wave}
  -\frac{1}{2}g_0^{cd}\partial^2_{X^cX^d}g^1_{ab} = 0,
\end{equation}
which is a wave-equation in the $X$ co-ordinates, endowed with a constant metric.

Suppose that the initial conditions for the metric (i.e. the initial tensor perturbations) are given by a the power-spectrum in equation \eqref{eqn:Power spectrum}; then initially, $\partial_X g_{1} = 0$, whence $\partial_X g_{1} = 0$ always.\footnote{Again, see \S \ref{subsec: consistency conditions stars} for an alternative derivation of the constancy of such a function.}


\subsection{Second-Order Equations}

The second-order terms in equations \eqref{eqn:wave operator} and \eqref{eqn:first differential operator} yield (using $\partial_X g_0 = \partial_X g_1 = 0$)
\begin{equation}
    -\frac{1}{2}g^{cd}_0\partial^2_{x^cx^d}g^0_{ab} - \frac{1}{2}g^{cd}_0\partial^2_{X^cX^d}g^2_{ab} + P_{ab}(g_0)\partial_x g_0 \partial_x g_0 - \Lambda g^0_{ab} = 8\pi G \rho^0_{ab}.
\end{equation}
Rearranging gives:
\begin{equation}\label{eqn:second order in epsilon}
  \begin{aligned}
    & \left[-\frac{1}{2}g^{cd}_0\partial^2_{x^cx^d}g^0_{ab} + P_{ab}(g_0)\partial_xg_0 \partial_xg_0 - \Lambda g^0_{ab}\right] \\ &
    -\frac{1}{2}\left[g^{cd}_0\partial^2_{X^cX^d}g^2_{ab} \right] = 8\pi G\rho^0_{ab}.
  \end{aligned}
\end{equation}
The first line is nothing but the Einstein tensor (and the $\Lambda$-term) for large scales. The other -- an oscillating part (with non-trivial initial conditions), that would vanish upon averaging, which includes the term $g^{cd}_0\partial^2_{X^cX^d}g^2_{ab}$, that dictates the evolution of the second-order perturbation of the metric. I consider this term in \S \ref{subsec: oscillatory}.

\subsection{The Averaged Part}
\label{subsec:averaged part}
Indeed, equation \eqref{eqn:second order in epsilon} may be broken into two parts: an averaged part (integrated, so to speak, over $X$), and an oscillating part. An advantage of the multiple-scales method is that averaging is only performed over a flat space-time, as opposed to other approaches to the averaging problem \cite{BuchertRaesaenen2012}. There are co-ordinates $\tilde{X}$ on $F_x$ in which $g^0_{ab}$ is the Minkowski metric (these co-ordinates depend on $x$, of course, but this is innocuous; see also appendix \ref{appendix: Newtonian}). In this co-ordinate system one may introduce a Fourier transform, which is carried out solely in a flat space-time, and is therefore unambiguous; then the average, $\langle f \rangle$, is simply the $\tilde{k} = 0$ component of the $\tilde{X}$-Fourier transform of $f$ (divided by the 4-volume). The Jacobian, $\sqrt{-\det g_0}$ is a constant, which is removed upon division by the $4$-volume. The oscillating part of $f$ is then $\{f\}_{\textrm{osc}} = f - \langle f \rangle$.

I shall show below that the oscillatory part of equation \eqref{eqn:second order in epsilon} may be solved consistently, leaving
\begin{equation}
  -\frac{1}{2}g^{cd}_0\partial^2_{x^cx^d}g^0_{ab} + P_{ab}(g_0)\partial_xg_0 \partial_xg_0 - \Lambda g^0_{ab} = 8\pi G\langle\rho^0_{ab}\rangle.
\end{equation}
This is with an Einstein equation in the $x$ co-ordinates for $g_0(x)$ -- a large-scale equation, sourced only by the averaged part of the energy-momentum tensor. This implies that, if $\rho_{ab} = O(1)$, then the small scales do not react back on the large-scale metric, to leading order. The exact functional form of $\langle \rho^0_{ab}\rangle$ may be guessed from symmetries -- from the cosmological principle -- to yield that $g_0(x)$ is an FLRW metric.

\subsection{The Oscillatory Part}
\label{subsec: oscillatory}
The oscillating part of equation \eqref{eqn:second order in epsilon} reads
\begin{equation}\label{eqn:oscillating}
  - \frac{1}{2}g^{cd}_0\partial^2_{X^cX^d}g^2_{ab} = 8\pi G \{\rho_{ab}^0\}_{\textrm{osc}}.
\end{equation}
This equation is a partial differential equation for $g_2$ -- a wave equation with a source.

By the existence and uniqueness theorem for the wave equation in flat space-time, this has a solution for any $\{\rho_{ab}^0\}_{\textrm{osc}}$; but my concern is to show that this solution does not break the asymptotic series, i.e. that the $g_2$ thus obtained does not become too big ($O(\eps^{-2})$).
Let me perform a Fourier transform in $\tilde{X}$.
The problem arises only from the resonant part of the energy-momentum tensor -- from its components that satisfy $\tilde{k}\cdot\tilde{k} = 0$, i.e. from relativistic motion on small scales. The other Fourier components of $\{\rho_{ab}^0\}_{\textrm{osc}}$ are chiefly non-relativistic matter particles, such as dark matter or stars, for whom $\tilde{k^0}^2 \gg \abs{\tilde{\mathbf{k}}}^2$. There is negligible contribution to the overall energy density from small-scale relativistic particles, but let us consider it anyway.
Indeed, by linearity, write $g^2 = g^2_{\textrm{non-rel}} + g^2_{\textrm{rel}} + g^2_{\textrm{hom}}$, and let each one of the first two summands be the solution to the wave equation, sourced by the non-relativistic and the relativistic parts of $\{\rho_{ab}^0\}_{\textrm{osc}}$ respectively, with zero initial conditions. The third satisfies a homogeneous wave-equation, and is so-far undetermined: it is
\begin{equation}
  g^2_{\textrm{hom},ab}(x,X) = \frac{1}{(2\pi)^4}\int \mathrm{d}^4\tilde{k} e^{-\mathrm{i}\tilde{k}\cdot\tilde{X}}\hat{g}^2_{\textrm{hom},ab}(x,\tilde{k})\delta(\tilde{k}\cdot\tilde{k}),
\end{equation}
where $a\cdot b = g^0_{cd}a^cb^d$.
This solution does not increase its amplitude, and therefore $\eps^{2}g^{2}_{\textrm{hom}}$ remains $O(\eps^{2})$, thus maintaining consistency. 
By the same argument, $g^2_{\textrm{non-rel}}$ maintains an amplitude that remains $O(1)$ throughout its evolution, which leaves only $g^2_{\textrm{rel}}$ as a potential problem.

Suppose that $\rho_{ab}$ contained a plane wave term $\exp(i\tilde{k}\cdot\tilde{X})$, where $\tilde{k}\cdot\tilde{k} = 0$. This would resonate with the wave-equation differential operator, producing a growing amplitude. If it became too large, there is a possibility that $\eps^2 g_2$ would grow larger than $\eps g_1$, thus ruining the asymptotic expansion. Plane waves due to the small-scale modes of cosmic microwave background are negligible thanks to diffusion damping \cite{Weinberg2008}. $8\pi G\rho_{ab}$ due to galaxy-scale electro-magnetic fields and neutrinos is assumed to be so small, that the resonant behaviour of the amplitude of $g^2_{ab}$ induced by it does not violate the asymptotic expansion (recall from \S \ref{subsec:fibre bundles} that the $X$-space-time is bounded).

If $\eps$ pertains to galactic scales of $\sim 1$ kpc, then the scales of coherent, relativistic motion of other particles tend to be much smaller, so that there are numerous such spatio-temporally confined resonant sources for $g^2_{\textrm{rel}}$. The associated Fourier components would, in general, have different phases, so, in effect, this contribution to $g^2_{\textrm{rel}}$ is the sum of waves emanating for point-like sources, with random phases. To find what $g^2_{\textrm{rel}}$ at each point $X$ is, one needs to superpose all the waves, each weighted by its source's distance from $X$. This problem has been considered in the past by refs. \cite{MeiksinWhite2003,Vincent2014,Ginatetal2019}, and the upshot is that, if the number of sources is finite, then the probability of $g^2_{\textrm{rel}}$ being higher than $h$ is $\sim h^{-3}$, for large $h$. Thus,
\begin{equation}
  P(g^2_{\textrm{rel}} \gtrsim \eps^{-p}) \propto N_{\textrm{tot}}\eps^{3p}.
\end{equation}
Therefore, $g_2$ is small in all probability (more rigorously, the asymptotic series may only converge in probability, but this is not a problem). Needless to say, the mean number of relativistic sources inside a galaxy is finite.

$g^2_{\textrm{hom}}$ is determined by imposing the consistency condition for the next order correction, $\eps^3g^3$. As $g^2_{\textrm{hom}}$ contains modes that are only relativistic, they resonate with the wave-equation operator. Imposing the consistency condition for $g^3$ results in a differential equation satisfied by $\hat{g}^2_{\textrm{hom},ab}(x,k)$ for all $k\neq 0$ (cf. \S \ref{subsec: consistency conditions stars} below). This equation determines $g^2_\textrm{hom}$ uniquely, using the initial conditions that are prescribed by the power-spectrum.

\section{Newtonian Objects}
\label{sec:Newtonian objects}
So far, I have explained why the leading order metric is unaffected by back-reaction caused by the small-scale oscillations of the energy-momentum tensor, \emph{as long as their amplitude is up to order unity}. In fact, it turned out to be completely independent of the small scale.

But usually, when one considers the averaging problem, one has the putative effect of over-densities $\delta\rho/\rho \gg 1$ in mind, which are typically present inside galaxies, primarily in stars. This would lead to an asymptotic series for $\rho_{ab}$ that includes terms of order, say, $\eps^{-2}$, which would change the low-order equations, leading to non-zero $X$ derivatives in the low-order terms in the asymptotic series of the metric $g$. Such terms could, conceivably, affect the $O(\eps^2)$ equation, which, as was shown earlier, dictates the large-scale behaviour of $g_0$.

My primary concern in this paper is to show that the technique I present here can be used to address this issue, and, given specific initial data (as well as a $\rho_{ab}$), to determine the extent to which the small-scale physics reacts back on the large-scale metric as well as the relevant consistency conditions. To do so I endeavour to find which terms in the $O(\eps^2)$ equation arise due to small-scale effects, below.

To make things less cumbersome, I include in $g_0$ any possible terms that are larger than $\ord{1}$. The equation for $g_0(\cdot,X)$ now reads
\begin{equation}\label{eqn:zeroth order stars}
  -\frac{1}{2}g^{cd}_0\partial^2_{X^cX^d}g^0_{ab} + P_{ab}(g_0)\partial_Xg_0\partial_Xg_0 = 8\pi G \rho^{-2}_{ab}.
\end{equation}
This equation is an Einstein equation with zero cosmological constant, whose sources are basically Newtonian point particles; the initial conditions are independent of $X$. As remarked in \S \ref{subsec:asymptotic expansion}, it is not assumed that $g_0$ is independent of $X$; rather, this will be demonstrated presently.

I assume that $\rho^{-2}_{ab}$ is due solely to Newtonian objects, which, as is well-known, generate, by themselves, a metric which is only a perturbation relative to flat space-time. This implies that in solving equation \eqref{eqn:zeroth order stars} one obtains two terms: $g_0(x)$, which describes the large-scale variation, and a perturbation, $h_{ab}(x,X)$, due to the stars. Its magnitude is of order $GM/R$ for a star,\footnote{Strictly speaking, this is correct in Newtonian gauge, relative a Minkowski background in the $\tilde{X}$ co-ordinates for $X$-space-time, defined as in \S \ref{subsec: oscillatory}.} which is about $\lambda \equiv 10^{-6} = 0.1\eps$; even for a galactic potential with circular rotation velocity of a few hundred kilometres per second, the magnitude of $h$ does not exceed this amount. Therefore, $h_{ab} = O(\eps)$ (in fact, $O(\lambda) \leq O(\eps)$), and may be safely absorbed into $g_1$, at the cost of its $X$-dependent parts increasing to $O(\lambda/\eps)$. See appendix \ref{appendix: Newtonian} for a more detailed calculation.
Additional contributions to $\rho_{ab}$ due to those Newtonian objects come from peculiar velocities, but these are of lower orders: terms in $T_{ab}$ like $\rho v_iv_j$ would be of order $\ord{\lambda^2\eps^{-2}} \sim \ord{\eps^0}$, while $0i$ terms like $\rho v_i$ would also enter at the level of $g^2$ only, so I ignore them here. Please see \S \ref{subsec:virialised objects} for more about them.

The second-order equation acquires two additional source terms, and reads
\begin{equation}\label{eqn:second order stars}
  \begin{aligned}
    & \left[-\frac{1}{2}g^{cd}_0\partial^2_{x^cx^d}g^0_{ab} + P_{ab}(g_0)\partial_xg_0 \partial_xg_0 - \Lambda g^0_{ab}\right] \\ &
    + \left[- g^{cd}_0\partial^2_{x^cX^d}g^1_{ab}- \frac{1}{2}g^{cd}_0\partial^2_{X^cX^d}g^2_{ab} + P_{ab}(g_0)\left(\partial_X g_1\partial_x g_0 + \partial_x g_0 \partial_X g_1\right)\right] \\ &
    + \left\{-\frac{1}{2}g^{cd}_1\partial^2_{X^cX^d}g^1_{ab} + P_{ab}(g_0)\partial_X g_1 \partial_X g_1\right\}_{\textrm{osc}} \\ &
    + \left\langle-\frac{1}{2}g^{cd}_1\partial^2_{X^cX^d}g^1_{ab} + P_{ab}(g_0)\partial_X g_1 \partial_X g_1\right\rangle = 8\pi G\rho^0_{ab}.
  \end{aligned}
\end{equation}

In \S \ref{subsec: consistency conditions stars} below, I show that one can solve the oscillatory part of the above equation consistently. What one is left with is
\begin{equation}\label{eqn:second order large scale stars}
  -\frac{1}{2}g^{cd}_0\partial^2_{x^cx^d}g^0_{ab} + P_{ab}(g_0)\partial_xg_0 \partial_xg_0 - \Lambda g^0_{ab} -B_{ab} = 8\pi G\langle \rho^0_{ab}\rangle,
\end{equation}
where
\begin{equation}\label{eqn:back-reaction tensor}
  B_{ab} = \left\langle\frac{1}{2}g^{cd}_1\partial^2_{X^cX^d}g^1_{ab} - P_{ab}(g_0)\partial_X g_1 \partial_X g_1\right\rangle.
\end{equation}
The tensor (on the $x$-space-time) $B_{ab}$ may constitute a possible back-reaction of the Newtonian sources, propagated through the non-linearity of the Einstein equations, on the large-scale properties of $g_0$ -- finding these was the goal of this section. Equations \eqref{eqn:second order large scale stars} and \eqref{eqn:back-reaction tensor} constitute something akin to a homogenised equation for cosmological back-reaction -- they describe the dynamics of the large-scale (leading-order) metric, taking small-scale inhomogeneities into account in a consistent manner. Recall that the averaging $\langle \cdot \rangle$ is carried out only in the flat fibre $F_x$, and is hence unambiguous. Equation \eqref{eqn:second order large scale stars} appeared already in the literature (e.g. \cite{Zalaletdinov1997,GreenWald2011}), but was derived differently. Here, instead of adopting \emph{a-priori} assumptions on the existence of a background metric and on the magnitude of deviations from it, the consistency conditions, which will be discussed in some length in \S \ref{subsec: consistency conditions stars}, allow one to quantify when these assumptions hold. In particular, the independence of $g_0$ from $X$ is derived above (see \S \ref{subsec: consistency conditions at low orders} for how the consistency condition implies that $\rho_\textrm{rel} = 0$), and the magnitude of the perturbations $\eps g_1 + \ldots$ is explicitly controlled by imposing that there are no resonant terms.

Appendix \ref{appendix: Newtonian} implies that
the $O(\lambda/\eps)$ term in $g^1_{ab}$, that corresponds to $h_{ab}$, is given, in the frame of reference of a freely-falling observer on $M$, by $\tilde{\zeta}_{ab}$, which is defined there. This frame is associated with an orthonormal tetrad $e^a_{\underline{b}}(x)$, which is used to convert from abstract indices to concrete ones (and \emph{vice versa}).
In this frame, the components $B_{\underline{ab}}$ of $B_{ab}$ are given by the expression in equation \eqref{eqn:back-reaction tensor}, with the derivatives in the $\tilde{X}$ system, and $g^1$ set to $\tilde{\zeta}$, \emph{viz.}
\begin{equation}\label{eqn:back-reaction tensor in freely-falling frame}
  \eps^2B_{\underline{ab}} = \left\langle\frac{1}{2}\tilde{\zeta}^{cd}\partial^2_{\tilde{X}^c\tilde{X}^d}\tilde{\zeta}_{\underline{ab}} - P_{\underline{ab}}(g_0)\partial_{\tilde{X}}\tilde{\zeta}\partial_{\tilde{X}}\tilde{\zeta}\right\rangle.
\end{equation}
The reason is, that even though $g_1$ and $g_0$ in equation \eqref{eqn:back-reaction tensor} are tensors on $M$ (and thus scalars on $F_x$), one actually performs two co-ordinate transformations here: one, on $M$, from harmonic co-ordinates to those of a freely-falling observer in $g_0$, at $x$, and then, an additional transformation on $F_x$ that takes $X$ to $\tilde{X}$. The average (as the zero mode of a Fourier transform) is invariant under the latter, which implies that the cumulative effect of both transformations justifies equation \eqref{eqn:back-reaction tensor in freely-falling frame}. Appendix \ref{appendix: gauge} explains how to calculate the components of $B_{ab}$ in a given gauge (which is not necessarily harmonic), provided that equation \eqref{eqn:second order large scale stars} is taken as an effective Einstein-like equation for $M$, and that one does not perform co-ordinate transformations whose derivative matrix is not $\ord{1}$.

Taking equation \eqref{eqn:second order large scale stars} as an Einstein equation, with $R_{ab}$ replacing $R^{(h)}_{ab}$, and with $B_{ab}$ now calculable in any gauge, one may perform a $3+1$ splitting and derive, \emph{inter alia}, a Raychaudhuri equation; the simplest way to do so is to move $B_{ab}$ to the matter side of the Einstein equation, and consider it as a correction to the energy-momentum tensor.

\section{Consistency Conditions}
\label{subsec: consistency conditions stars}
The two middle rows of equation \eqref{eqn:second order stars} are important for ensuring the consistency of the asymptotic expansion $g \sim g^0 + \eps g^1 + \ldots$: they constitute the updated equation \eqref{eqn:oscillating}, with $g_1$ having an additional $O(\lambda/\eps)$ component that is due to Newtonian sources. The latter would not engender resonances precisely because it arises from non-relativistic objects. In any case, together with $\{\rho_{ab}^0\}_{\textrm{osc}}$, these average out to zero on large scales. These terms will be addressed in \S \ref{subsec:consistency second order stars} below. First I wish to start with a more general discussion of consistency conditions at low order (order $1$), and to show what happens if the consistency conditions are not satisfied -- as in the case of relativistic sources that dominate the energy-momentum tensor at order $-1$. Assuming such sources do not exist, I return to the second-order consistency condition in \S \ref{subsec:consistency second order stars}, and show explicitly that it is satisfied under the assumptions of \S \ref{sec:Newtonian objects}. Then I comment on consistency conditions at higher orders, and on the effect of gravitational waves on those conditions.

\subsection{Consistency Conditions at Low Orders}
\label{subsec: consistency conditions at low orders}
Let us start by examining the consistency conditions at low orders, before moving on to second order. Including a possible relativistic contribution to $\rho^{-1}_{ab}$, $\rho^{-1}_{\textrm{rel},ab}$, the first order equation reads (please recall, that by the discussion following equation \eqref{eqn:zeroth order stars}, $8\pi G\rho^{-2}_{ab}$ is actually $O(\lambda\eps^{-2}) \sim O(\eps^{-1})$)
\begin{equation}\label{eqn:first order stars relativistic}
  8\pi G(\rho^{-2}_{ab} + \rho^{-1}_{\textrm{rel},ab}) = -\frac{1}{2}g_0^{cd}\partial^2_{X^cX^d}g^1_{ab}.
\end{equation}
This is a linear equation, so it might be written as
\begin{equation}
  g^1_{ab} = h_{ab}/\eps + g^1_{\textrm{rel},ab} + g^1_{\textrm{hom},ab},
\end{equation}
where $h_{ab}$ is the same as in \S \ref{sec:Newtonian objects} above -- i.e. sourced only by the Newtonian objects comprising $\rho^{-2}_{ab}$ with zero initial conditions, $g^1_\textrm{rel}$ is sourced by $\rho^{-1}_\textrm{rel}$, and $g^1_{\textrm{hom}}$ is a solution of the homogeneous equation, to be determined by initial conditions. Unfortunately, due to the fact that $\partial_X g^0 = 0$, there is no freedom left to impose a consistency condition, which will remove the resonant contribution $g^1_\textrm{rel}$. Thus, one is forced to conclude that $g^1$ grows due to resonances, and breaks the asymptotic expansion. Hence, unless $\rho^{-1}_\textrm{rel} = 0$, the asymptotic expansion in this paper is inadequate. To clarify: the consistency condition at $\ord{\eps}$ is $\rho^{-1}_\textrm{rel} = 0$.

If the asymptotic expansion does break down, this implies, unsurprisingly, that the background $g_0$ cannot be independent of the small scale. While indeed equation \eqref{eqn:zeroth order} still holds for $g^0_{ab}$, it is no longer true that the initial conditions are $X$-independent. Indeed, if $\rho^{-1}_\textrm{rel} \neq 0$, the above discussion implies an order zero $X$-dependent contribution to them, stemming from the resonant term. In that case, one would have, instead of equation \eqref{eqn:first order stars relativistic},
\begin{equation}
\begin{aligned}
  8\pi G(\rho^{-2}_{ab} + \rho^{-1}_{\textrm{rel},ab}) & = -\frac{1}{2}g_0^{cd}\partial^2_{X^cX^d}g^1_{ab} - g_0^{cd}\partial^2_{x^cX^d}g^0_{ab} \\ &
  - \frac{1}{2}g_1^{cd}\partial^2_{X^cX^d}g^0_{ab} + g_1^{pq}\frac{\partial P_ab}{\partial g^{pq}}\partial_X g_0 \partial_X g_0 + 2P_{ab}\partial_x g_0 \partial_X g_0.
\end{aligned}
\end{equation}
Now the extra $X$-dependence of $g_0$ implies that one may impose the revised consistency condition
\begin{equation}
  8\pi G \hat{\rho}^{-1}_{\textrm{rel},ab}(x,k) = \int \mathrm{d}^4 \tilde{X}~e^{\mathrm{i}\tilde{k}\cdot\tilde{X}}\left[- g_0^{cd}\partial^2_{x^cX^d}g^0_{ab} + 2P_{ab}\partial_x g_0 \partial_X g_0 \right],
\end{equation}
which would ensure that $g^1$ does remain small, relative to $g^0$, at the expense of introducing an $X$-dependence to $g^0$. However, this extra complication arises only if, contrary to the assumptions made in this paper, there are relativistic sources at orders more dominant than $o(\eps^0)$. The purpose of this discussion was to show what happens when the consistency conditions break down, and what their break-down's implications are. Carrying on with the assumption of Newtonian sources, the first non-trivial consistency condition arises only at second order.

\subsection{Second Order}
\label{subsec:consistency second order stars}
Returning to equation \eqref{eqn:second order stars}: as shown above, $g^1$ contains a homogeneous term $g^1_{\textrm{hom},ab}$, of the form
\begin{equation}
  g^1_{\textrm{hom},ab}(x,X) = \frac{1}{(2\pi)^4}\int \mathrm{d}^4\tilde{k} e^{-\mathrm{i}\tilde{k}\cdot\tilde{X}}\hat{g}^1_{\textrm{hom},ab}(x,\tilde{k})\delta(\tilde{k}\cdot\tilde{k}),
\end{equation}
which is, as yet, undetermined.\footnote{In \S \ref{subsec:low-order equations} I ignored this term, arguing that $g^1$ is independent of $X$. Indeed, if I had introduced it, then from the consistency condition \eqref{eqn:consistency for g1}, one would have been able to deduce that $\partial_X g_1 = 0$. The same goes for $\partial_X g_0$.} As $g^1$ has no relativistic sources, this term is the only one with modes that satisfy $\tilde{k}\cdot \tilde{k} = 0$.

Imposing the consistency condition, i.e. that the source terms for $g^2(\cdot,X)$ not contain resonant parts, constrains this homogeneous part. Explicitly, for all $k\neq 0$ such that $\tilde{k}\cdot\tilde{k} = 0$,
\begin{equation}\label{eqn:consistency for g1}
\begin{aligned}
  & -k^c\partial_{x^c}\hat{g}^1_{\textrm{hom},ab}(x,k) + P_{ab}^{cdefgh}(g_0)\left(k_c\hat{g}^1_{\textrm{hom},ef}(x,k)\partial_{x^d}g^0_{gh} + k_d\hat{g}^1_{\textrm{hom},gh}(x,k)\partial_{x^c}g^0_{ef}\right) \\ &
  - \mathrm{i}\int \mathrm{d}^4\tilde{X}e^{\mathrm{i}\tilde{k}\cdot\tilde{X}}\left\{ P_{ab}(g_0)\partial_X g_1 \partial_X g_1  - \frac{1}{2}g^{cd}_1\partial^2_{X^cX^d}g^1_{ab}\right\}_{\textrm{osc}} = -8\pi \mathrm{i}G \hat{\rho}^0_{ab}(x,k),
\end{aligned}
\end{equation}
where $\hat{\rho}^0_{ab}(x,k)$ is the contribution to $\rho^0_{ab}$ from null wave-vectors.

Equation \eqref{eqn:consistency for g1} is a differential equation that determines the free part $\hat{g}^1_{\textrm{hom}}$, given initial data (in $x$), except for the averaged part with $k=0$, $\hat{g}^1_{\textrm{hom}}(x,0)$. The latter is determined by the average (over $X$) of the $\ord{\eps^3}$ equation. The oscillating part of the $\ord{\eps^3}$ equation serves to determine $g^3(\cdot,X)$, as well as the consistency condition for $g^2$. Let us see that in more detail: as explained in \S \ref{subsec: oscillatory} above, although $\hat{\rho}^0_{ab}(x,k) = 0$, for null $\tilde{k}$ -- since electro-magnetic fields at small scales are expected to be negligible at this order and likewise any CMB contribution -- this will not be the case in general at higher orders. All other contributions to $\{\rho^0_{ab}(x,X)\}_{\textrm{osc}}$ are not null, i.e. non-resonant, and therefore do not appear in equation \eqref{eqn:consistency for g1}.

The readers should bear in mind, that so far, $g^1_{\textrm{hom}}$ is undetermined, and I have not used any initial conditions for $g^1_{ab}$. These initial conditions are set at one time $t_\textrm{rec}$, and so cannot be used immediately in equation \eqref{eqn:first order stars relativistic}, except at $x$ such that $x^0 = t_\textrm{rec}$.\footnote{More generally they may be specified on Cauchy surface, not necessarily on a surface of constant $x^0$, but this complication does not make a difference.} Equation \eqref{eqn:consistency for g1} is a differential equation for the tensor-valued function $\hat{g}^1_{\textrm{hom}}(x,k)$ on $T^{*\otimes2}M$ (not on $T^{*\otimes2}F_x$, as one may initially expect for a small-scale equation), and it will determine $g^1_\textrm{hom}$ from the initial date, such that it is ensured that the consistency conditions are satisfied, so that the asymptotic expansion does not break down. The homogeneous part is not constrained by any other equation at this order. It is a quasi-linear first order partial differential equation -- only the first term on the right is a derivative term, and as such may be solved for any (sufficiently smooth) initial data \citep[appendix IV, \S 4]{Choquet-Bruhat2009}.

The power-spectrum \eqref{eqn:Power spectrum} implies that these initial data -- for $g^1$ -- are zero. Furthermore,
\begin{equation}
\int \mathrm{d}^4\tilde{X}e^{\mathrm{i}\tilde{k}\cdot\tilde{X}}\left\{ P_{ab}(g_0)\partial_X (g^1 - g^1_\textrm{hom}) \partial_X (g^1 - g^1_\textrm{hom})  - \frac{1}{2\eps^2}h^{cd}\partial^2_{X^cX^d}h_{ab}\right\}_{\tilde{k}^2 = 0} = 0
\end{equation}
because a product of two non-relativistic modes, with $\tilde{k}^0 \gg |\tilde{\mathbf{k}}|$, does not contain a non-zero null mode, whence equation \eqref{eqn:consistency for g1} implies that $\hat{g}^1_\textrm{hom}(x,k) = 0$ for $\tilde{k} \neq 0$. We will see in the next sub-section how the \emph{third} order equation determines $\hat{g}^1_\textrm{hom}(x,0)$.


\subsection{Third Order}
Now $g^1_{ab}$ is determined completely, save for $\hat{g}^1_{\textrm{hom},ab}(x,k=0)$. This term is governed by the third order Einstein equation, which reads
\begin{equation}\label{eqn:third order stars}
\begin{aligned}
  8\pi G \rho^1_{ab} & = -\frac{1}{2}\left[g_0^{cd}\partial^2_{X^cX^d}g^3_{ab} + g_1^{cd}\partial^2_{X^cX^d}g^2_{ab} + g_2^{cd}\partial^2_{X^cX^d}g^1_{ab}\right] \\ &
  -\left[g_0^{cd}\partial^2_{x^cX^d}g^2_{ab} + 2g_1^{cd}\partial^2_{x^cX^d}g^1_{ab}\right]
  -\frac{1}{2}\left[g_0^{cd}\partial^2_{x^cx^d}g^1_{ab} + g_1^{cd}\partial^2_{x^cx^d}g^0_{ab}\right] \\ &
  + P_{ab}(g_0)\left[\partial_X g_1\partial_X g_2 + \partial_X g_2\partial_X g_1 + \partial_x g_1\partial_X g_1 + \partial_X g_1\partial_x g_1 + \partial_X g_2\partial_x g_0 + \partial_x g_0\partial_X g_2\right] \\ &
  + P_{ab}(g_0)\left[\partial_x g_1 \partial_xg_0 + \partial_x g_0 \partial_xg_1\right] + \Lambda g^1_{ab}\\ &
  + g^1_{pq}\frac{\partial P_{ab}}{\partial g_{pq}}(g_0)\left[\partial_x g_0 \partial_xg_0 + \partial_X g_1 \partial_xg_0 + \partial_x g_0 \partial_Xg_1 + \partial_X g_1 \partial_Xg_1\right].
\end{aligned}
\end{equation}
A few observations about this equation are in order: at third order, $g_0$ is completely determined, $g_1$, as remarked, is fully determined, except for a large-scale homogeneous term, $g_2$ is determined except for a homogeneous solution of the wave-equation, and $g_3$ is fully unknown. As was the case in lower orders, this equation splits into a small-scale equation, which determines the small-scale behaviour of $g^3_{ab}$ -- up to a homogeneous solution (via the operator $-\frac{1}{2}g_0^{cd}\partial^2_{X^cX^d}g^3_{ab}$), and a large-scale one, which would actually give $\hat{g}^1_{\textrm{hom},ab}(x,0)$ and which contains a back-reaction term, too. The former equation also gives the consistency condition at this order, which would constrain $g^2_\textrm{hom}$. 

The averaged part of equation \eqref{eqn:third order stars} is
\begin{equation}\label{eqn:third order large scale stars}
\begin{aligned}
  8\pi G \langle \rho^1_{ab}\rangle & = -\frac{1}{2}\left[g_0^{cd}\partial^2_{x^cx^d}\langle g^1_{ab}\rangle + \langle g_1^{cd}\rangle\partial^2_{x^cx^d}g^0_{ab}\right] + P_{ab}(g_0)\left[\partial_x \langle g_1 \rangle \partial_xg_0 + \partial_x g_0 \partial_x\langle g_1\rangle\right] + \Lambda\langle g^1_{ab}\rangle \\ &
  -\frac{1}{2}\left\langle g_1^{cd}\partial^2_{X^cX^d}g^2_{ab} + g_2^{cd}\partial^2_{X^cX^d}g^1_{ab} + 4g_1^{cd}\partial^2_{x^cX^d}g^1_{ab}\right\rangle
  + \langle g^1_{pq}\rangle\frac{\partial P_{ab}}{\partial g_{pq}}(g_0)\partial_{x} g_{0} \partial_{x}g_0 \\ &
  +P_{ab}(g_0)\left\langle\partial_X g_1\partial_X g_2 + \partial_X g_2\partial_X g_1 + \partial_x g_1\partial_X g_1 + \partial_X g_1\partial_x g_1\right\rangle \\ &
  + \frac{\partial P_{ab}^{cdefgh}}{\partial g_{pq}}(g_0)\left\langle g^1_{pq}\partial_{X^c}g^1_{ef} \partial_{x^d}g^0_{gh} + g^1_{pq}\partial_{x^c} g^0_{ef} \partial_{X^d}g^1_{gh} + g^1_{pq}\partial_{X^c} g^1_{ef} \partial_{X^d}g^1_{gh}\right\rangle,
\end{aligned}
\end{equation}
and its last three lines constitute a back-reaction effect. The unknown in this equation is $\hat{g}^1_{\textrm{hom},ab}(x,0)$, and upon inspection one realises that equation \eqref{eqn:third order large scale stars} is actually linear in it, which implies the existence and uniqueness of a solution in the usual manner. Observe, that while the above discussion on the consistency condition \eqref{eqn:consistency for g1} precludes high-frequency gravitational radiation at $\ord{\eps}$, it does allow for low-frequency gravitational waves at this order, through $\hat{g}^1_\textrm{hom}(x,0)$. This is entirely consistent with the power-spectrum in equation \eqref{eqn:Power spectrum}, which implies that primordial high-frequency gravitational waves begin at $\ord{\eps^2}$.

Now let us proceed to the oscillating equation, and thence derive the consistency condition for $g_2$. As before, the latter consists of the requirement that the $\tilde{k}\cdot\tilde{k} = 0$ terms vanish, which implies that, for such $k$:
\begin{equation}\label{eqn:consistency for g2}
\begin{aligned}
  8\pi G \hat{\rho}^1_{ab} & = -\mathrm{i}k^c\partial_{x^c}\hat{g}^2_{ab} + \mathrm{i}P_{ab}^{cdefgh}(g_0)\left[k_c\hat{g}^2_{ed}\partial_{x^d} g^0_{gh} + \partial_{x^c} g^0_{ef}k_d\hat{g}^2_{gh}\right] \\ &
  -\frac{1}{2}\left[g_0^{cd}\partial^2_{x^cx^d}\hat{g}^1_{ab} + \hat{g}_1^{cd}\partial^2_{x^cx^d}g^0_{ab}\right]
  + P_{ab}(g_0)\left[\partial_x \hat{g}_1 \partial_xg_0 + \partial_x g_0 \partial_x\hat{g}_1\right] + \Lambda \hat{g}^1_{ab}\\ &
  + \int \mathrm{d}^4\tilde{X}~e^{\mathrm{i}\tilde{k}\cdot\tilde{X}}\bigg\{-\frac{1}{2}\left[g_1^{cd}\partial^2_{X^cX^d}g^2_{ab} + g_2^{cd}\partial^2_{X^cX^d}g^1_{ab} + 4g_1^{cd}\partial^2_{x^cX^d}g^1_{ab}\right] \\ &
  + P_{ab}(g_0)\left[\partial_X g_1\partial_X g_2 + \partial_X g_2\partial_X g_1 + \partial_x g_1\partial_X g_1 + \partial_X g_1\partial_x g_1\right] \\ &
  + g^1_{pq}\frac{\partial P_{ab}}{\partial g_{pq}}(g_0)\left[\partial_x g_0 \partial_xg_0 + \partial_X g_1 \partial_xg_0 + \partial_x g_0 \partial_Xg_1 + \partial_X g_1 \partial_Xg_1\right]\bigg\}_{\textrm{osc}}.
\end{aligned}
\end{equation}
As before, this should be solved to determine the homogeneous piece $g^2_\textrm{hom}$. While the consistency condition for $g^1_\textrm{hom}$ was a non-linear equation (which was not a problem since $\hat{g}^1_\textrm{hom}(x,k)$ vanished due to the initial data), it was unique in being so, and indeed equation \eqref{eqn:consistency for g2} is linear in $\hat{g}^2(x,k)$. Thus, it can be solved, and its growth is controlled by energy estimates \citep[appendix IV, \S 4]{Choquet-Bruhat2009}. There are two main differences with respect to the previous consistency condition: firstly, while previously the relativistic contribution to $\rho^0_{ab}$ was zero, there is no such restriction here; secondly, as discussed below equation \eqref{eqn:Power spectrum}, the initial conditions for $\hat{g}^2_{ab}(x,k)$ and for $\partial_x \hat{g}^2_{ab}$ are non-vanishing. The is crucial if one wants to be able to account for gravitational waves. The linearity of equation \eqref{eqn:consistency for g2} -- in contrast to the non-linearity of equation \eqref{eqn:consistency for g1} -- implies that neither non-vanishing initial conditions nor gravitational waves or a non-vanishing relativistic contribution to $\hat{\rho}^1_{ab}$ pose any difficulty.

It might be the case that equation \eqref{eqn:consistency for g2} implies that $g^2_\textrm{hom}$ grows indefinitely in $x^0$. Let me give a qualitative argument why this would still not break the asymptotic expansion: if $g_0$ varies on a (large) scale $\lambda_0$, and $g_1(x,\cdot)$ -- on a scale $\lambda_1$ (this pertains to $x$, not $X$), and if $P_{ab}(g_0)$ and its derivatives are $\ord{1}$, then equation \eqref{eqn:consistency for g2} implies, roughly, that
\begin{equation}
\begin{aligned}
  k^c \partial_{x^c}g_2 & \sim 8\pi G \rho^1 + \Lambda g_1 + k\lambda_0 g_2g_0 + g_0g_1(\lambda_1^2 + \lambda_0^2 + 2\lambda_1\lambda_0) \\ &
  + k^2g_1g_2 + k\lambda_1g_1^2 + \lambda_0^2 g_1g_0^2 + k\lambda_0g_0g_1^2 + k^2g_1^3,
\end{aligned}
\end{equation}
whence $\hat{g}^2_\textrm{hom}$ varies on a scale $\lambda_2$ which is set by $k,\lambda_0,\lambda_1$, all of which are of order unity (recall that $k$ is the conjugate of $X$, which is re-scaled to have the same units as $x$); so $\lambda_2$ must also be of the same order. Therefore, if, indeed, $g^2_\textrm{hom}$ grows to such a magnitude that it dominates over all source terms in the averaged equation \eqref{eqn:third order large scale stars}, then the latter equation would imply that
\begin{equation}
  g_1g_0\left(\lambda_1^{-2} + \lambda_0^{-2}\right) \sim g_1 g_2\left(\lambda_2^{-2} + \lambda_1^{-2} + \lambda_1^{-1}\lambda_2^{-1}\right).
\end{equation}
This implies that $g_0 \sim g_2$, whence, by appendix \ref{appendix: Newtonian}, $g_1 \sim g_2$ as well; the hierarchy $g_0 \gg \eps g_1 \gg \eps^2 g_2$ is therefore still maintained. Having, discussed the $\ord{\eps^3}$ condition, let me move on to sketch how the next order equations look like in the next sub-section.

\subsection{Higher Order -- A Sketch}
This situation repeats itself at higher orders: the equation at $\ord{\eps^n}$ is, qualitatively,
\begin{equation}\label{eqn: nth order qualitative}
  -\frac{1}{2}g^{cd}_0\partial^2_{X^cX^d}g^n_{ab} + \left(\partial_X ~\textrm{terms}\right) + \left(X ~\textrm{indep.}\right) + \Lambda g^{n-2}_{ab}= 8\pi G\rho^{n-2}_{ab},
\end{equation}
where the second term only involves terms that contain derivatives $\partial_X$, acting on $g^k$, with $k < n$, and are not independent of $X$. The $X$-independent part contains the zero mode of the former terms, as well as terms involving $g^0$ only (if there are any). The first two terms on the left-hand-side, as well as the oscillatory part of $\rho^{n-2}_{ab}$ determine the consistency condition on $\hat{g}^{n-1}_{ab}(x,k)$, which fixes it uniquely using the relevant initial data, as well as the $X$-dependence on $g^n$ (up to a homogeneous solution of the wave equation, which will be determined by the consistency condition at the next order and initial data). This consistency condition is, like equation \eqref{eqn:consistency for g2}, linear in $\hat{g}^{n-1}(x,k)$. The averaged part of equation \eqref{eqn: nth order qualitative} determines $\hat{g}^{n-2}(x,k=0)$.

To summarise this section, the consistency conditions emerge naturally from the requirement that the asymptotic expansion which was used to solve the Einstein equations remain asymptotic. That is, that $\eps^n g^n \gg \eps^{n+1}g^{n+1}$. Analysing them, it turned out that the requirement could be satisfied if the following conditions obtained:
\begin{enumerate}
  \item There are no small-scale relativistic sources up to (and including) $\rho^0_{ab}$, so that
  \begin{equation}
    \hat{\rho}^{-2}_{ab}(x,k) = \hat{\rho}^{-1}_{ab}(x,k) = \hat{\rho}^{0}_{ab}(x,k) = 0;
  \end{equation}
  and
  \item the initial data are such that any $X$-dependence begins at $\ord{\eps^2}$.
\end{enumerate}
If these obtain, the asymptotic expansion is valid; otherwise, it breaks down. If all sources below $\ord{1}$ are Newtonian, but the initial power-spectrum is different, then a multiple-scales solution is still possible, but a different expansion is necessary (e.g. in powers of $\sqrt{\eps}$). If, on the other hand, the first of the two conditions above fails, we have seen in \S \ref{subsec: consistency conditions at low orders} that this implies that the `background' $g_0$ ceases to be independent of the small scale.

Having discussed the consistency conditions, which are necessary for equation \eqref{eqn:second order large scale stars} to be valid, one may move on to study some examples.

\section{Applications}
\label{sec:applications}
To solve equation \eqref{eqn:second order large scale stars}, one needs to know $B_{ab}$. But it depends on $g^1$, whose $x$-behaviour is governed by the $\ord{\eps^3}$ equation. One can, however, solve it iteratively, by expanding in the magnitude $b$ of $B_{ab}$, while at the same time going to higher orders in $\eps$, as needed.


To zeroth order in $b$, one solves equation \eqref{eqn:second order large scale stars}, to get $g^0$. Then one should substitute it into the third order equation, i.e. the $x$-equation for $g^1$, which would give $g^1(x,X)$, while ignoring the back-reaction terms \emph{there} (it is of higher order in $b$); this would have to be combined with a solution of the (non-linear) matter equations of motion, to obtain $\rho^{-2}$. Solving both would yield $g^1_{ab}(x,X)$.\footnote{Recall that throughout this paper, I assumed that $\rho_{ab}$ was given, but to apply this technique in a practical setting -- a simulation for instance -- one would have to find $\rho_{ab}$, too. Please see appendix \ref{appendix:filtering of rho} for more on how to obtain $\rho_{ab}(x,X)$ from $\rho_{ab}(x,x/\eps)$ found in, e.g., a simulation.} Inserting this into the definition of $B_{ab}$ and solving equation \eqref{eqn:second order large scale stars} once again, one finds a corrected solution for $g^0$, accurate to order $0$ in $\eps$, and first order in $b$.

One can go further and solve the fourth-order equation to find $g^2$, compute the back-reaction (to order $b$) on $g^1$, and then correct $B_{ab}$ for $g^0$, giving a solution for $g^0$ to second order in $b$, and $g^1$ to first order in $b$. Combining the two yields a solution $g^0 + \eps g^1$, which is correct to $\ord{\eps}$ and $\ord{b}$. This procedure could be continued up to the required accuracy.

Suppose that one has obtained a solution to first order in $\eps$, $g_{ab} = g^0_{ab}(x) + \eps g^1_{ab}(x,X)$. To compare it with an exact solution, or use it in practice, one must substitute $X = x/\eps$ back in, because the treatment of $X$ and $x$ as independent is only a mathematical technique, designed to derive the solution. Once obtained, it must describe a four-dimensional space-time.

\subsection{Scope and Error Term}
Let me briefly discuss the error on equation \eqref{eqn:second order large scale stars}. As seen in \S \ref{subsec: consistency conditions stars}, there are non-trivial consistency conditions that must be satisfied for the whole asymptotic expansion to be valid. One further condition, for the small-large-scale splitting, was that the only perturbations with amplitude $\gg \ord{1}$ to $\rho_{ab}$ be on small (order $\eps$) scales.
Provided these obtain, the error on using the solution $g^0_{ab}(x)$ of equation \eqref{eqn:second order large scale stars} instead of the exact solution $g_{ab}$ of equation \eqref{eqn:Einstein} is of course given by
\begin{equation}
  \delta g_{ab} = \eps g^1_{ab}\left(x,\frac{x}{\eps}\right) + \ldots,
\end{equation}
at leading order.

I have shown above that one sufficient condition for this error to be small is that all matter in the Universe be non-relativistic (in the sense of the Newtonian approximation holding \emph{locally}). Indeed, if one also includes relativistic, gravitating objects such as black holes, the technique for solving equation \eqref{eqn:zeroth order stars} -- by writing the solution as $g^0_{ab}(x) + \eps g^1_{ab}(x,X)$, where $g^1$ contains all of the $X$-dependence -- fails. At present I do not know how to surmount this obstacle.

\subsection{Examples}
\label{subsec:examples}
In this sub-section I compute the back-reaction tensor explicitly in a few examples. These are models where the metric is known, and is used to compute $B_{ab}$ (in contrast with the complementary case, described at the beginning of \S \ref{sec:applications}, where the metric is not known). The first is an example of a small-scale over-density -- a halo -- which might yield some back-reaction. The other two are cases where the technique presented in this paper is used to re-derive previously-known results: that virialised objects do not induce back-reaction \citep{Baumannetal2012} and that there is no back-reaction in a `Swiss-cheese'-type model, where a Schwarzschild metric inside some sphere is matched to an Einstein-de-Sitter universe.

\subsubsection{NFW Halo}
\label{subsec:NFW}
Consider, for instance, a galaxy with a constant-in-time Navarro-Frenk-White \cite{NFW1997,Binney} profile, embedded in some background universe. The halo's potential is
\begin{equation}
  \Phi(r) = -4\pi G\rho_0r_0^2\frac{\ln(1+r/r_0)}{r/r_0} \equiv -4\pi G\rho_0r_0^2f(y),
\end{equation}
with $r_{200} = 15r_0$, $M = 10^{12} ~M_\odot = 200\rho_{\textrm{crit}}\frac{4}{3}\pi r_{200}^3$, $y = r/r_0$, where $\rho_{\textrm{crit}} = 27.75\times 10^{10}h^2 ~M_\odot ~\textrm{Mpc}^{-3}$, and $H_0 = 67.4 ~\textrm{km s}^{-1}~\textrm{Mpc}^{-1} = 100h ~\textrm{km s}^{-1}~\textrm{Mpc}^{-1}$ \citep{Planck2018}. Locally (in $x$), one can use the results of appendix \ref{appendix: Newtonian} to compute the back-reaction tensor.
One can impose the asymptotic expansion $g \sim g^0(x) + \eps g^1(x,X) + \ldots$, requiring that $g^0$ be independent of $X$. Thus, denoting by $g^0$ the underlying metric (so-far undetermined), and by $\eps g^1$ the rest, one obtains an expansion which is unambiguous to $\ord{\eps}$. The back-reaction tensor may now be easily calculated using equation \eqref{eqn:back-reaction tensor}.

Take, for example, $a = b = i$ for some Cartesian index. Then, by equation \eqref{eqn:P_abcdefgh}
\begin{equation}
  B_{\underline{ii}} = \left\langle\frac{\partial_i g^1_{ii}\partial_i g_1^{ii}}{2} + \textrm{Christoffel product}+ \textrm{boundary term}\right\rangle,
\end{equation}
where I have integrated the second derivative by parts once (the boundary term turns out to be minuscule).
Hence (the integral $\mathrm{d}X^0$ cancels out with the $X^0$ dimension of the $4$-volume) each of the terms consists of a product of two first derivatives of $\Phi$, each of which is about
\begin{align}
  \left\langle\partial_X g_1 \partial_X g_1\right\rangle \sim \eps^{-2}\times \frac{G^2\rho_0^2r_0^4}{c^4}\frac{3(4\pi)^3}{4\pi (r_{200}/r_0)^3}\int_{0}^{r_{200}/r_0}f'(y)^2y^2\mathrm{d}y \approx 6.6\times 10^{-5} ~(100 ~\textrm{Mpc})^{-2}
\end{align}
Of course, the units of $100$ Mpc are those in which $x$ is expressed. This is small (a few percent) compared with $\frac{8\pi G \rho_{\textrm{crit}}}{c^2} \approx 0.0015 ~(100 ~\textrm{Mpc})^{-2}$, whence it emerges that the back-reaction due to averaging exists in this model, but is small relative to the background. This makes the anzats of setting $g^0$ to be independent of $X$ self-consistent.

In fact, the potential needs to be truncated outside $r_{200}$, beyond which point it falls off like $1/r$, i.e.
\begin{equation}
  \Phi = \begin{cases}
           -4\pi G\rho_0r_0^2f(y) + 4\pi G\rho_0r_0^2f(15) - \frac{GM}{15r_0}, & \mbox{if } y \leq 15 \\
           -\frac{GM}{r_0y}, & \mbox{otherwise},
         \end{cases}
\end{equation}
where the constants have been added such that $\Phi(y) \underset{y\to\infty}{\to} 0$. It is clear, however, that the value of $\left\langle\partial_X g_1 \partial_X g_1\right\rangle$ computed above, which corresponds to $B_{ab}(x^i = 0)$, serves as an upper bound on $B_{ab}(x)$. The latter is obtained, of course, by averaging $\partial \Phi \partial \Phi$ over spheres of radius $r_0$, centred at $x$.

This calculation carries forward directly to the case of many such haloes at random different positions $x$, provided that the haloes do not overlap. If they did, this would constitute a term in $\rho_{ab}$ that is $\gg O(1)$ over a scale which is much longer than the one set by $\eps$, thereby violating the assumption made in the first paragraph of \S \ref{sec:small-scale}. If the haloes are distributed randomly in space, and the typical separation between haloes is much smaller than $100 ~\textrm{Mpc}$ -- the scale of $x$, and if it is larger than the haloes' size, one can deduce that, on the large scale $x$, the back-reaction tensor $B_{ab}$ should be homogeneous and isotropic. Together with $\langle \rho_{ab}^0\rangle$, this implies that, in this particular case, $g^0_{ab}$ is close to an FLRW metric.

\subsubsection{General Virialised Objects}
\label{subsec:virialised objects}

One further observation is now in order: if one defines the matrix $U_{ab} = \langle \partial_{\tilde{X}^a}\Phi\partial_{\tilde{X}^b}\Phi\rangle$, then, up to negligible boundary terms, $B_{\underline{ab}}$ may be expressed in terms of $U_{ab}$, which is related to the average density of (Newtonian) gravitational potential energy. Indeed, for a virialised system, where $\frac{1}{c}\abs{\frac{\partial \Phi}{\partial \tilde{X}^0}} \ll \abs{\tilde{\del}\Phi}$, $U_{ab}$ is precisely the matrix version of the density of gravitational potential energy in Newtonian gravity. Thus, roughly speaking, at least for Newtonian sources for $\rho^{-2}_{ab}$, the back-reaction tensor $B_{ab}$ represents a correction to the Einstein equations, in which small-scale gravitational potential energy, averaged over the small scale, also gravitates.
In this sub-section, as an example of the application of the technique developed in this paper, I wish to reproduce the result of ref. \cite[\S 4.5]{Baumannetal2012}, that a virialised Newtonian object on an FLRW background does not induce any back-reaction. The derivation here proceeds by first relating $B_{ab}$ to the Newtonian gravitational potential energy, and then the kinetic part of $\langle \rho^0_{ab}\rangle$ -- to the Newtonian kinetic energy. In a freely-falling frame in $g^0_{ab}$, I take the 3-velocity field of the particles to be $v_i(\tilde{X})$, and the density field as $\rho(\tilde{X})$. In the section only, I omit the $\sim$, when it is not confusing to do so.

Using equation \eqref{eqn:back-reaction tensor in freely-falling frame}, one may associate $B_{\underline{ab}}$ with a second order expansion of the Ricci tensor over a flat background, in harmonic gauge. Working out all the terms and imposing the gauge condition yields eventually that
\begin{equation}
  \eps^2B_{\underline{ab}} = \left\langle\frac{1}{4}\partial_{\tilde{X}^{\underline{a}}}\tilde{\zeta}_{cd}\partial_{\tilde{X}^{\underline{b}}}\tilde{\zeta}^{cd} - \frac{1}{4}\partial_{\tilde{X}^{e}}\tilde{\zeta}\partial_{\tilde{X}^{\underline{a}}}\tilde{\zeta}^e_{\underline{b}} \right\rangle,
\end{equation}
whence, in a perturbed Newtonian metric as in appendix \ref{appendix: Newtonian}, $\eps^2 B_{\underline{ij}} = 2U_{ij}$.

Let us recall the definition of the Chandrasekhar potential energy matrix from Newtonian gravity (see, e.g., \cite{Binney}),
\begin{equation}
  \eps^2 W_{ij} \equiv -\int \rho X_i\frac{\partial \Phi}{\partial X^j} \mathrm{d}^3X = \frac{1}{4\pi G}\int \frac{\partial \Phi}{\partial X^i}\frac{\partial \Phi}{\partial X^j}\mathrm{d}^3X - \frac{\delta_{ij}}{8\pi G}\int \frac{\partial \Phi}{\partial X^k}\frac{\partial \Phi}{\partial X^l}\delta^{kl}\mathrm{d}^3X.
\end{equation}
This matrix can be related to $B_{ab}$ via $B_{\underline{ij}} = 8\pi G(W_{ij} - \delta_{ij}W)$, whence the trace-revered version of $B_{ab}$ is related to $W_{ij}$ via
\begin{equation}\label{eqn:virial B and W}
  B_{\underline{ij}} - \frac{1}{2}\eta_{\underline{ij}}B = 8\pi G W_{ij},
\end{equation}
where $\eta_{\underline{ab}}$ is the Minkowski metric (corresponding to $g^0_{ab}$ in the current reference frame). This is the potential energy part.

As for the kinetic energy: as mentioned in \S \ref{sec:Newtonian objects} above, $\rho^{-2}_{ab}$ is only due to Newtonian objects, which implies that it is dominated by its $00$ component. The $ij$ components are $\ord{\lambda^2\eps^{-2}} \sim O(1)$, which means that they enter into $\rho^0_{ab}$; i.e. $\langle \rho^0_{\underline{ij}}\rangle$ contains the (trace-reverse of) $\langle \rho v_iv_j\rangle$ of the same Newtonian objects which gave rise to $\rho^{-2}_{ab}$, in addition to the energy-momentum tensor of the background FLRW fluid.\footnote{The $0i$ components of the energy-momentum tensor of those Newtonian objects is $\ord{\lambda/\eps^2}$, i.e. order $-1$, but its average is zero. Its contribution to $\delta g$ is down-scaled by another factor of $\lambda$, which means that it enters only in $g^2$, but this is harmless.} Its trace-reversed version may therefore be decomposed as
\begin{equation}\label{eqn:virial T and K}
  \langle T^0_{\underline{ij}} \rangle \equiv \left\langle \rho^0_{\underline{ij}} - \frac{1}{2}\eta_{\underline{ij}}\rho^0\right\rangle = T_{\underline{ij}}^{FLRW} + \langle\rho v_iv_j\rangle + \mbox{h.o.t.} = T_{\underline{ij}}^{FLRW} + 2K_{ij},
\end{equation}
where $K_{ij}$ is the Newtonian kinetic energy matrix.

The trace-reversed version of equation \eqref{eqn:second order large scale stars} is
\begin{equation}
  R_{\underline{ab}} - \frac{1}{2}R\eta_{\underline{ab}} + \Lambda \eta_{\underline{ab}} = B_{\underline{ab}} - \frac{1}{2}B\eta_{\underline{ab}} + 8\pi G \langle T^0_{\underline{ab}}\rangle,
\end{equation}
where the quantities on the left-hand side are computed with $g^0$, and the derivatives are $x$-derivatives. Equations \eqref{eqn:virial B and W} and \eqref{eqn:virial T and K} imply that the right-hand side of the $\underline{ij}$ component is equal to
\begin{equation}
  T_{\underline{ij}}^{FLRW} + 8\pi G(W_{ij} + 2K_{ij}).
\end{equation}
If the virial theorem holds, $W_{ij} + 2K_{ij} = 0$, yielding an Einstein equation which is solved by an FLRW metric.\footnote{Similar results are valid for the $00$ and $0i$ components, by imposing the normalisation condition of the 4-velocity field, $u_{\underline{a}}u^{\underline{a}} = -1$.} This agrees with the results of \cite{Baumannetal2012}. The derivation proceeded in an observer frame, so it may be generalised to backgrounds different from FLRW (or a perturbed FLRW space-time).

\subsubsection{Swiss-Cheese-Like Model}
Finally, let me consider, very briefly, one additional known example of (the absence of) back-reaction. Consider an Einstein-de-Sitter space-time (an $\Omega_m = 1$, FLRW space-time, with $a(t) \propto t^{2/3}$), in which a sphere of radius $R = a(t)\eps$ is cut around the spatial origin. The matter inside the sphere is collapsed to a star (not a black hole, as this would violate some of the assumptions made here, but this variation is innocuous). By Birkhoff's theorem this space-time is described by a Tolman-Oppenheimer-Volkoff metric inside the star, a Schwarzschild metric outside the star, until $R$, and a dust FLRW metric outside $R$, all of which can be matched continuously \cite{EinsteinStraus1945}.

By construction this space-time has no back-reaction, in-so-far-as the metric far from the star is always FLRW. Indeed, if one takes $\eps$ to be the small scale (I ignore the case $a \ll 1$ or $a \gg 1$, but these can also be handled in the same way), then it is immediate from equation \eqref{eqn:back-reaction tensor in freely-falling frame} that $B_{ab}$ is non-zero \emph{only} at $r = 0$, where it is finite. This is a removable discontinuity, and thus does not influence the metric away from the origin; this fact meshes well with the expected result of no back-reaction.



\section{Discussion}
\label{sec:discussion}

In this paper I presented an approach to study the averaging problem in cosmology using the method of multiple scales. The small and the large scales were treated as independent variables in harmonic gauge, and the Einstein field equations were expanded in the small scale. This yielded perturbative equations for both the small and the large scales, which were solved iteratively, up to second order in $\eps$; at this order one obtained an effective equation for the large-scale dependence of the metric, which also includes a back-reaction term, and a consistency condition for the asymptotic expansion, which was shown to be satisfied in general. I studied the consistency conditions, which played a major part in the multiple-scales analysis, at orders up to $\ord{\eps^3}$ and demonstrated how they determine the homogeneous parts of the wave-equation solutions. Solutions to the consistency-condition equations were shown to exist under the conditions in \S \ref{sec:Newtonian objects}. 

Let me re-emphasise the importance of the consistency conditions: if they are na\"{i}vely ignored, then all terms, starting from $g^1_{ab}$, are simply undetermined. Alternatively, if they are not satisfied, then the entire asymptotic expansion breaks. Conversely, though, solving them and ensuring that solutions exist ensures the validity of the expansion, and renders the separation of the metric to an $X$-independent background $g_0$ and perturbations to it meaningful. These conditions are therefore a \emph{sine qua non} in any multiple-scales analysis of the averaging problem.

I also showed that the back-reaction term vanishes completely if the energy-momentum tensor is always of the same order as the averaged one (at most), but it does not in general. If the $O(\eps^{-2})$ density variations are due to Newtonian objects, then the back-reaction terms are small. However, a detailed model of the small-scale system is needed to study back-reaction in a realistic system \cite{Korzynski2015}.

Throughout the paper I assumed that the effect of black holes and neutron stars may be neglected. While this might be justified by Birkhoff's theorem for isolated bodies (if the distance between them and the next over-density is $\gg GM$), it cannot be used to treat the fully relativistic case. I have shown that the asymptotic expansion in \S \ref{sec:small-scale} does not break down up to second order, but higher orders may be achieved as discussed in \S \ref{sec:applications}. One can generalise the approach I presented here to account for these issues, as well as to derive bounds on the error. On the other hand, this approach has the advantage that it does not require any averaging over curved manifolds, and is effective in revealing the terms in the Einstein equations that lead to possible back-reaction, and how to gauge their magnitude.

Due to the well-separation of stellar, galactic and large-scale-structure scales, one can extend the formalism presented in this paper to account for back-reaction due to inhomogeneities on all of these scales, by introducing $X_{\textrm{stars}}$, $X_{\textrm{gal}}$, $X_{\textrm{LSS}}$, in addition to the cosmological-scale $x$, and treating all four variables as independent.

The approach I proposed here and its relation to the averaging problem, are quite analogous to Hamiltonian perturbation theory, when faced with a resonance (say, in the context of celestial mechanics). Usually, the equations of motion are obtained there by averaging over the fast variables -- the mean anomalies of the individual bodies (analogous to small scales) \cite{Arnoldetal2006} -- thereby generating averaged equations of motion which govern the evolution of the slow variables (such as the energies and angular momenta). But na\"{i}ve averaging cannot be done when a resonance is present, which is simply another way of saying that the fast variables react back on the slow ones. Instead, resonant perturbation theory is required, which also draws on the method of multiple scales.


\acknowledgments{I wish to thank Vincent Desjacques, Robert Reischke, Fabian Schmidt, Robert Lilow, Stefano Anselmi and Dennis Stock for helpful discussions and comments. I acknowledge funding from the Israel Science Foundation (grant no. 1395/16 and 255/18) and from the Israeli Academy of Sciences' Adams Fellowship.}

\bibliographystyle{JHEP}
\bibliography{averaging}

\appendix
\section{Explicit Solution With Newtonian Objects}
\label{appendix: Newtonian}
The purpose of this appendix is the calculation of $h_{ab}$ is \S \ref{sec:Newtonian objects}. Equation \eqref{eqn:zeroth order stars} consists of a metric with constant (in $X$) boundary conditions, with an energy-momentum tensor that is comprised of Newtonian masses. This equation, \emph{qua} a partial differential equation, is an Einstein equation (written in harmonic co-ordinates), which describes what one would like to identify with a metric on $F_x$. Making this identification is akin to studying the tensor $g_F \in T^{*\otimes2}\bund$, given by
\begin{equation}
  g_F = g_{ab}(x,X)\frac{\partial}{\partial X^a}\otimes \frac{\partial}{\partial X^b}.
\end{equation}
(As before this is actually in a sub-space corresponding to $T^{*\otimes2}F_x$.)
This tensor is not to be confused with the metric $g^0_M = g^0_{ab}$ on $M$, although they have the same components.

As the energy-momentum tensor is generated by Newtonian sources, one can solve this equation perturbatively, writing $g_F = g_F^0 + g_F^1 = g^0_{ab} + \zeta_{ab}$, where $g^0_{ab}$ is a function of $x$ only (i.e. a constant on $F_x$), and $\zeta_{ab} = \ord{\lambda}$. As in \S \ref{subsec:averaged part}, one transforms to a co-ordinate system $\tilde{X}$ on $F_x$ where $\tilde{g}_F^0 = \eta$ is the Minkowski metric; the transformation is $\tilde{X}^b = P_a^bX^a$ (it exists due to Sylvester's law of inertia).
The transformation matrix\footnote{I assume that this matrix is $\ord{1}$ in $\eps$.} $P_a^b = P_a^b(x)$ has the same components as the matrix that transforms $g_M$ to the co-ordinates of a freely-falling observer on $M$ with metric $g_0$ (although, as before, the former lies in the tangent space of $F_x$ whereas the latter -- in the tangent space of $M$), i.e. $\eta_{ab}P^a_cP^b_d = g^0_{cd}$. This transformation leaves $\zeta_{ab}$ in harmonic gauge. However, Newtonian gauge is harmonic for particles whose (peculiar) velocities are much lower than the speed of light, so in this approximation I take $\zeta_{ab}$ to be in Newtonian gauge, whence it emerges that
\begin{align}
  & \tilde{\zeta}_{00} = -2\Phi \\&
  \tilde{\zeta}_{ij} = -2\Phi\delta_{ij}\\ &
  \tilde{\zeta}_{0i} = 0,
\end{align}
where $\Phi$ is the Newtonian gravitational potential.


\section{Gauge Transformations}
\label{appendix: gauge}

Having computed $B_{\underline{ab}}$ in a frame attached to a freely-falling observer on $M$ in $g_0$, and having seen that it is small relative to $\rho_\textrm{crit}$, one may wish to calculate $B_{ab}$ in a specific co-ordinate system; for instance, in conformal Newtonian gauge. To see how this is done, I ignore any large-scale perturbations to the density field (and to the velocity field). Then to zeroth order in $b$ (see \S \ref{sec:applications}), $g_0$ is the FLRW metric, and a freely-falling observer there is co-moving.

The observer's tetrad may be taken as $e^a_{\underline{0}} = u^a$, and $e^b_{\underline{i}} = \delta^b_{\underline{i}}/a(\eta)\sqrt{\gamma_{ii}}$ (no sum is implied), where $a(\eta)$ is the scale-factor and the spatial part of the metric is $g^0_{ij} = a^2\gamma_{ij}$. Then
\begin{equation}\label{eqn:Back-reactio tensor in some gauge}
  B^{ab} = B^{\underline{cd}}e^a_{\underline{c}}e^b_{\underline{d}};
\end{equation}
explicitly (no sum implied),
\begin{equation}\label{eqn:B_components}
\begin{aligned}
  B^{00} & = \frac{B^{\underline{00}}}{a^2} \\
  B^{0i} & = \frac{B^{\underline{0i}}}{a^2\sqrt{\gamma_{ii}}} \\
  B^{ij} & = \frac{B^{\underline{ij}}}{a^2\sqrt{\gamma_{ii}\gamma_{jj}}}.
\end{aligned}
\end{equation}

If there exist large-scale perturbations, then one has to perform an additional asymptotic expansion in both the magnitude of these perturbations, and $\eps$. Re-summing the former would imply that all the large-scale perturbations are present in the $g_0$ of this paper; the procedure for obtaining $B_{ab}$ in this case is the same as was outlined above, \emph{mutatis mutandis}. To first order in this re-summation in Newtonian gauge, the metric is a perturbed FLRW metric, given by
\begin{equation}
  \mathrm{d}s_0^2 = a^2\left[-(1+2\Phi)\mathrm{d}\eta^2 + (1+2\Psi)\delta_{ij}\mathrm{d}x^i\mathrm{d}x^j\right],
\end{equation}
which means that $u^a = a^{-1}(1-\Phi,\mathbf{v}_{\textrm{pec}})$, and $e^a_{\underline{i}}$ are chosen to make the tetrad orthonormal. Equation \eqref{eqn:Back-reactio tensor in some gauge} still holds, of course. Besides, a perturbed FLRW metric is necessary in general, because even if there are no large-scale perturbations, $B_{ab}$ itself would produce them (to first order in $b$, and higher). Equations \eqref{eqn:B_components} are already at first order in the large-scale perturbation, so they receive contributions from the change of $u^a$ only at second order.

\section{Filtering of the Energy-Momentum Tensor}
\label{appendix:filtering of rho}
In this appendix I present one way to derive $\rho_{ab}(x,X)$ from $\rho_{ab}(x,x/\eps)$, where the latter might arise from, e.g., a numerical simulation. As the simulation is performed in a 4-dimensional space-time, its solution would be a tensor-field $\rho_{ab}(x,x/\eps)$, which might not have an explicit dependence on $\eps$, so a procedure is needed to derive the $X$ dependence of $\rho_{ab}$ from it, in a systematic manner.
The simplest way to do so is to use a co-ordinate \emph{dependent} Fourier transform. Please bear in mind that the Fourier transform discussed in this appendix is \emph{not} the same as the fibre Fourier transform that is described in the main body of the paper; the purpose of what is described here is \emph{only} to explain how one might arrive at the starting point of \S \ref{sec:small-scale}, which is a given tensor-field $\rho_{ab}(x,X)$.

Working in harmonic co-ordinates, one might use the manifold's charts to define a matrix $\rho^\pb_{ab}\in \mathbb{R}^{4\times4}$, by pulling $\rho_{ab}$ back: if $p$ is a point in space-time, and $\psi:\mathbb{R}^4\to M$ is a chart associated with harmonic co-ordinates $x$, then $\rho^\pb_{ab}(x) = \psi^*(\rho_{ab}(p))$, where $p = \psi(x)$. As a matrix-valued real function, one might use the usual Fourier transform on $\mathbb{R}^4$, and define
\begin{align}
  & \hat{\rho}_{ab}^\pb(k) = \int e^{\mathrm{i}k\cdot x}\rho^\pb_{ab}(x) \mathrm{d}^4x \\ &
  \rho_{ab}^\pb(x) = \frac{1}{(2\pi)^4}\int e^{-\mathrm{i}k\cdot x}\hat{\rho}^\pb_{ab}(k) \mathrm{d}^4k.
\end{align}
The filtering may now proceed in the usual way, by splitting the integral in the second line into two parts: with $\abs{k} > 2\pi/\eps$, and with $\abs{k}\leq 2\pi/\eps$ (the absolute values are computed in the Euclidean metric on $\mathbb{R}^4$). Then, the second integral is used to define the $X$-dependent part of $\rho_{ab}$:
\begin{equation}\label{eqn:filtering of rho}
  \rho_{ab}^\pb(x,X) = \frac{1}{(2\pi)^4}\int_{\abs{k} \leq 2\pi/\eps} e^{-\mathrm{i}k\cdot x}\hat{\rho}^\pb_{ab}(k) \mathrm{d}^4k + \frac{1}{(2\pi)^4}\int_{\abs{k} > 2\pi/\eps} e^{-\mathrm{i}k\cdot X\eps}\hat{\rho}^\pb_{ab}(k) \mathrm{d}^4k.
\end{equation}

Equation \eqref{eqn:filtering of rho} is not a unique way to obtain $\rho_{ab}(x,X)$ from $\rho_{ab}(x,x/\eps)$. Other ways are also possible, such as an expansion in eigenvectors of the Laplace-Beltrami operator on $M$, or a different transform in harmonic co-ordinates (such as a spherical-harmonics decomposition, related to equation \eqref{eqn:filtering of rho} by Rayleigh's identity). Each way of defining $\rho_{ab}(x,X)$ yields a (possibly) different asymptotic series \eqref{eqn:asymptotic expansion metric}. This reflects on the non-uniqueness of asymptotic expansions in general, and does not pose a problem. If the $\rho_{ab}(x,X)$ found in a different approach satisfies (i) $\rho_{ab} = \rho_{ab}(x,x/\eps)$ and (ii) all the conditions mentioned in the paper, in particular that the metric terms $g^1$, $g^2$, \emph{etc.} obtained as in \S \ref{sec:small-scale} and \S \ref{sec:Newtonian objects} satisfy the consistency conditions, then the asymptotic expansions, resulting from applying the method of multiple scales to the different approach and to equation \eqref{eqn:filtering of rho}, must be consistent with each other.

\end{document}